\documentclass{aa}
\usepackage{graphics,lscape,epsfig}

\def\la{\mathrel{\hbox{\rlap{\hbox{\lower4pt\hbox{$\sim$}}}\hbox{$<$}}}}
\def\ga{\mathrel{\hbox{\rlap{\hbox{\lower4pt\hbox{$\sim$}}}\hbox{$>$}}}}

\def\arcmin{\hbox{$^\prime$}}
\def\arcsec{\hbox{$^{\prime\prime}$}}

\def\fm{\hbox{$.\!\!^{\rm m}$}}

\def\fdg{\hbox{$.\!\!^\circ$}}
\def\farcm{\hbox{$.\mkern-4mu^\prime$}}
\def\farcs{\hbox{$.\!\!^{\prime\prime}$}}

\newcommand{\etal}{{et al.}\,}      
\def\deg{{^\circ}}

\newcommand{\msun}{\,{\rm M}_\odot}

\newcommand{\kms}{{\,km\,s$^{-1}$}}

\newcommand{\HI}{\mbox{\normalsize H\thinspace\footnotesize I}}

\begin{document}
\title{Parkes H\,I observations of galaxies behind the southern Milky Way.}
\thanks{Tables 1 and 2 are available in electronic format at the CDS
via anonymous ftp to cdsarc.u-strasbg.fr (130.79.128.5) or via
http://cdsweb.u-strasbg.fr/Abstract.html}

\subtitle{I. The Hydra/Antlia region ($l\approx$ 266$\deg$ to 296$\deg$)}

\author{Ren\'ee C. Kraan-Korteweg\inst{1} 
\and Patricia A. Henning\inst{2}
\and Anja C. Schr\"oder\inst{3}}

\offprints{Ren\'ee C. Kraan-Korteweg,
\email{kraan@astro.ugto.mx}}

\institute{
Depto.~de Astronom\'\i{a}, Universidad de Guanajuato, Apartado Postal 144, 
Guanajuato, Gto 36000, Mexico
\and
Institute for Astrophysics, University of New Mexico, 800 Yale Blvd., NE, 
Albuquerque, NM, 87131, USA
\and 
Dept. of Physics and Astronomy, University of Leicester, University Road,
Leicester LE1 7RH, U.K.}

\date{Received date;accepted date}

\titlerunning{\HI\ observations of galaxies in the southern ZOA. I.}
\authorrunning{R.C.~Kraan-Korteweg et al.}
  
 
\abstract{As part of our program to map the large-scale distribution
of galaxies behind the Milky Way, we used the Parkes 210\,ft (64\,m)
radio telescope for pointed \HI\ observations of a sample of low
surface-brightness (due to heavy obscuration) spiral galaxies selected
from the deep optical Zone of Avoidance (ZOA) galaxy catalog in the
Hydra/Antlia region (Kraan-Korteweg 2000a). Searching a simultaneous
velocity range of either 300 to 5500\kms\ or 300 to 10\,500\kms\ to
an rms level of typically $2 - 4$\,m\,Jy resulted in detections in 61
of the 139 pointings, leading to a total of 66 detections (an
additional detection was made in a reference position, and two other
pointings revealed two and four independent signals
respectively). Except for 2 strong \HI\ emitters identified in the
shallow Zone of Avoidance \HI\ survey (Henning et al. 2000), all
\HI\ detections are new. An analysis of the properties of the observed
and detected galaxies prove that pointed \HI\ observations of highly
obscured galaxies allow the tracing of a population of nearby,
intrinsically large and bright spiral galaxies that otherwise would
not be recovered. The new data identified a previously unrecognized
nearby group at $\ell \sim 287\fdg5, b \sim-9\fdg5, V\sim 1700$\kms,
the continuation of the Hydra/Antlia filament on the opposite side of
the Galactic plane, and helped to delimit a distinct void in the ZOA
centered at 2000\kms.
}
\maketitle

\section{Introduction}

Until recent years, galaxy surveyors have avoided the difficult
territory of the Milky Way.  Great strides toward filling in the
low-galactic-latitude gap in our knowledge of the galaxian
distribution have been made by optical, radio, near- and far-infrared,
and X-ray surveys (see review by Kraan-Korteweg \& Lahav 2000, and the
many contributions in ``Mapping the Hidden Universe'', ASP
Conf. Ser. 218, eds. Kraan-Korteweg \etal 2000).  The first step in
this cartography (with the exception of 21-cm surveys which
immediately yield the angular coordinates and redshift of galaxies) is
the two-dimensional mapping of galaxies.  Follow-up redshift
measurements are required to map the distribution of galaxies in three
dimensions.  The two-dimensional galaxy distribution alone can be
misleading.  For instance, the prominent overdensity of galaxies in
Vela ($\ell \sim 280\deg, b \sim +6\deg$) made apparent by the optical
search by Kraan-Korteweg (2000a) in the Hydra/Antlia extension was
found to be due to a superposition of a nearby ($\sim 3000$\kms)
filament connecting to the Hydra cluster, a more distant ($\sim
6000$\kms) shallow extended supercluster, and a very distant ($\sim
16\,000$\kms) wall-like structure crossing the ZOA (Kraan-Korteweg \etal
1995).
 
In this paper, we will present the results from pointed
\HI\ observations of a sample of low surface-brightness (LSB) obscured
spiral galaxies selected from the deep optical galaxy catalog in the
ZOA ($266\deg \la \ell \la 296\deg, -10\deg \la b \la +8\deg$) in the
extension of the Hydra and Antlia clusters (Kraan-Korteweg 2000a). The
optical catalog is the first of a series of five 
systematic galaxy searches covering the southern Milky Way between
Galactic longitudes $245\deg$ and $350\deg$ (see Fig.~1 in
Kraan-Korteweg 2000a). Galaxies were identified by eye through the
systematic inspection of 50-times amplified images of the IIIaJ film
copies of the ESO/SRC sky survey. Although the majority of the newly
uncovered galaxies are small and faint, many of them may be
intrinsically large and nearby but reduced in apparent size, magnitude and
classifiable morphology due to the increasing foreground dust
absorption at lower latitudes.

An analysis of the catalogs (see Kraan-Korteweg 2000a and Woudt \&
Kraan-Korteweg 2001 for details) show that galaxies above a diameter
limit of $D \ga 0\farcm2$ (determined at an isophote of approximately
24.5 mag/arcsec$^2$) can be discovered ``easily'' through obscuration
layers of 3 magnitudes of extinction. Up to that extinction level of
$A_{\rm B} = 3\fm0$, the catalogs are complete for galaxies with intrinsic
diameters of $D^{\rm o} \ge 1.3$~arcmin (extinction-corrected, i.e. the
diameter they would have if they were not lying behind the Milky
Way). The optical searches succeed in a reduction of the solid angle
of the ZOA by a factor of about $2 - 2.5$, respectively from $A_{\rm B} =
1\fm0$ to $A_{\rm B} = 3\fm0$ (see also Fig.~\ref{dist}).

In order to also reduce the gap in redshift space, redshifts are
required for a representative sample of this newly optically filled-in
part of the ZOA. For the determination of the peculiar velocity of the
Local Group and the mapping of the velocity flow field, the nearby
galaxy population is particularly important. We obtain individual
optical spectroscopy of the brightest galaxies with the SAAO 1.9\,m
telescope (Kraan-Korteweg \etal 1995 for the
Hydra/Antlia ZOA region; Fairall \etal 1998, and Woudt \etal 1999 for
the Crux and Great Attractor regions respectively) and 21-cm line
observations of extended LSB spirals with the 64m Parkes radio
telescope (this paper), and finally, low resolution, multifiber spectroscopy
for the high-density regions (Felenbok et al.~1997, Woudt \etal
2002). We typically measure recession velocities for about
15\% of our ZOA galaxies.

These three observing methods are complementary in
galaxy populations, characteristic magnitude, diameter range, and the
depth of volume they probe: whereas the multifiber spectroscopy gives
a good description of clusters and dense groups in the ZOA out to
recession velocities of 25\,000\,\kms, the SAAO and \HI\ observations
cover the bright end of the galaxy distribution and provide a fairly
homogeneous sampling of galaxies out to 10\,000\,\kms (Kraan-Korteweg
\etal 1994).

The \HI\ observations are vital in recovering an important fraction of
the nearby spiral galaxy population which would otherwise be
impossible to map.  In addition to the extremely obscured and/or LSB
galaxies, some of the brighter spiral galaxies with redshifts known
from optical spectroscopy were reobserved with the Parkes
radiotelescope, as their \HI\ data are relevant to our program of
mapping the peculiar velocity field in the ZOA via the Tully--Fisher
relation.

Our pointed \HI\ observations are complementary to the ongoing blind
\HI\ survey in the ZOA, conducted with the multibeam receiver on the
Parkes telescope (Henning \etal 2000, Staveley-Smith \etal 2000b).
Unlike the current work, the blind \HI\ ZOA survey is not optically
selected, but is rather a fully-sampled survey over the region
$212\deg \le \ell \le 36\deg$, $|b| \leq 5\deg$.  The velocity coverage
of the blind survey is $-1200$ to 12\,700\,\kms, and the sensitivities
obtained by the deep blind survey will be similar to those obtained in
the current work.  Note that most of our objects lie outside of the
blind \HI\ survey region.

In the following section, a description of the observations is
given. Sect.~3 then provides the \HI\ data and profiles of the detected
galaxies, including details about a number of pointings which revealed
more than one signal, two of which do not seem to have an optical
counterpart even though the extinction is not extreme at those
positions. In Sect.~4 the observed galaxies that were not detected in
\HI\ are listed with the searched velocity range. This is followed by
an analysis of the properties of the detected (and non-detected)
galaxies (Sect.~5) and a brief discussion of the resulting galaxy distribution
in redshift space in relation to known features adjacent to the ZOA (Sect.~6),
and finally, a summary is provided in Sect.~7.

\section{Observations} \label{obs}

The Parkes 64\,m radio telescope\footnote{The Parkes telescope is part
of the Australia Telescope which is funded by the Commonwealth of
Australia for operation as a National Facility managed by CSIRO.} was
used over four observing periods of about 10 -- 14 days each (June
1993, April 1994, July 1995 and September 1996). Here, we report on
the observations that cover the Hydra/Antlia ZOA region, which were
made mainly during the first 2 observing periods (see Fig.~\ref{dist}
for an outline of the search area).

All observations were carried out in total power mode. Ten minute
ON-source observations were preceded by an equal length OFF-source
observation at the same declination but 10.5 minutes earlier in right
ascension, so as to traverse the same path in topocentric coordinates
during both the reference and the signal observation. Such
ON/OFF-observations were typically repeated three times, less for
strong sources with a clear detection after a shorter integration
time, and more in a few cases when a weak possible detection was
identified after this sequence.  In those cases, observations were
repeated until the reality of the signal was unambiguous.

At 21 cm, the telescope has a half-power beamwidth (HPBW) of
15$\arcmin$, and the system temperature was typically 39~K at the time
of these observations.  In 1993, we used the 1024-channel
autocorrelator to cover a bandwidth of 32\,MHz, using the two
orthogonal linear polarizations.  With the bandwidth generally
centered at 3000\kms, we obtained good S/N coverage for the velocity
range $300 - 5500$\kms. In a second step, we reobserved some of the
non-detections at a central velocity of 7500\kms. With this set-up,
the channel spacing was 6.6\kms\ and the velocity resolution after
Hanning smoothing was 13.2\kms.  From 1994 on, we used two IF's and
offset 512 channels of each polarization by 22~MHz. This resulted in
an instantaneous velocity coverage of about $300 - 10\,500$\kms\, with a
channel spacing of 13.2\kms, and a velocity resolution after Hanning
smoothing of 27.0\kms.  Although the lower frequencies were often
badly disturbed by interference around 8300\kms, this increased the
average detection rate considerably.

The online control program automatically corrected for the zenith
angle dependence of the telescope sensitivity. The pointing and
primary flux calibration was obtained (by local personnel) from
observations of the continuum source Hydra A (Davies et al. 1989).  As
an added check of system performance, secondary \HI\ flux calibrators,
chosen from the General Catalog of \HI\ Observations by Huchtmeier \&
Richter (1989) were observed regularly in all the runs (mainly
NGC\,1232, IC\,5201, Anon\,0203-55, ESO\,501-23, as well as a few
others). Based on those data the internal consistency of the flux
scale is of the order of 15\%.

The data were reduced using the Spectral Line Analysis Program
(``SLAP''; Staveley-Smith 1985). In all cases the two orthogonal
polarizations were averaged during data reduction.  Usually a third-
to fifth-order polynomial baseline was subtracted from the spectra.
With a few exceptions for the 512 channel observations, the spectra
were then generally Hanning-smoothed once.

\section{Detections}

In the following, we present the parameters of the detected galaxies.
The reduced \HI\ spectra of the detected galaxies are shown in
Fig.~\ref{profile}.  The optical properties (observed and
extinction-corrected) as well as the \HI\ parameters are then given
Table~\ref{det}.  The columns in the Table are explained below.

{\bf Col. 1:} Identification as given in the optical Hydra/Antlia
ZOA galaxy catalog (Kraan-Korteweg 2000a). The superscript ``$^{\rm a}$''
indicates that the Parkes telescope was actually not pointed towards
the galaxy listed in Table~\ref{det} but towards another galaxy at
small angular separation (see also galaxies marked with ``$^{\rm a}$'' in
Table~\ref{ndet}). Based on independent optical redshift information,
the detected signal could be unambigously assigned to this galaxy
instead of the galaxy centered in the beam. The signal of the galaxy
RKK\,1288 was obtained from the pointing to RKK\,1284 at an angular
separation of $\Delta r = 9\farcm4$; RKK\,2546 from RKK\,2525 at
$\Delta r = 4\farcm3$ (see further details in Sect.~4.1). Obviously
the \HI\ fluxes of these detections will be slightly underestimated as
the measurements were obtained off-center. This is indicated by a
``*'' in the columns describing the \HI\ parameters (column 14 -- 17).

While disentangling these ambiguities it was noted that the optical
parameters of the galaxies RKK\,2544 and RKK\,2546 are interchanged in
the ZOA catalog and thus also in the SAAO redshift paper
(Kraan-Korteweg \etal 1995). It has been corrected in 
Table~\ref{det}.
  
{\bf Col. 2:} Second name. Most of the second identifications given
in Column 2 of Table~\ref{det} originate from the ESO/Uppsala Survey
of the ESO(B) Atlas (Lauberts, 1982), recognizable as 'L' plus the
respective field and running number. FGCE\# stands for the Flat Galaxy
Catalog (Karenchentsev \etal 1993), AM\# for Arp Madore (1987), and
HIZSS for the shallow \HI\ Parkes All Sky Survey (HIPASS) ZOA survey
behind the southern Milky Way (Henning et al. 2000).  In three of the
pointed observations (RKK1037, RKK1919 and RKK1947) more than one
21-cm detection was identified within the beam and covered velocity
range.  Some of these \HI\ detections do not have an optical
counterpart (INV\#). These special cases will be discussed in further
detail later in this section.

{\bf Col. 3:} Identification in the IRAS Point Source Catalog (IRAS PSC
hereafter).  The entries indicate I (certain identification), P
(possible identification), and Q (questionable). See Table~3 and
Sect.~5 in Kraan-Korteweg (2000a) for further details.

{\bf Col. 4:} Right Ascension RA (J2000.0).

{\bf Col. 5:} Declination Dec (J2000.0). The ':' signifies that the
coordinates have a lower accuracy compared to the 1 arcsec precision
of the galaxies in the ZOA catalog.

{\bf Col. 6:} Galactic longitude $\ell$.

{\bf Col. 7:} Galactic latitude $b$.

{\bf Col. 8:} Morphological type. The morphological types are coded
similarly to the precepts of the RC2 (de Vaucouleurs \etal 1976) with
the addition of the subtypes E, M and L.  They stand for early spiral
(S0/a-Sab), middle spiral (Sb-Sd) and late spiral or irregular
(Sdm-Im).

{\bf Col. 9:} Large diameter $D$ and small diameter $d$ in arcsec.

{\bf Col. 10:} Apparent magnitude $B_{\rm J}$. These magnitudes are
eye-estimates from the ESO/SERC IIIaJ film copies. They compare well
with the Lauberts \& Valentijn (1989) ${\rm B_{25}}$ magnitudes and
have a 1$\sigma$ dispersion of less than $0\fm5$.

{\bf Col. 11:} The Galactic reddening at the position of the galaxy,
as given by the DIRBE/IRAS extinction maps (Schlegel \etal 1998). See
the catalog paper for a more detailed discussion (Kraan-Korteweg 2000a).

{\bf Col. 12:} Extinction-corrected large diameter. The
corrections are based on Cameron's (1990) law and use the conversion
value to extinction in the blue, $A_{\rm B} = 4.14 \cdot E_{\rm (B-V)}$,
according to Cardelli \etal (1989).

{\bf Col. 13:} Extinction-corrected apparent magnitude, again
applying the laws as given in Cameron (1990).

{\bf Col. 14:} Heliocentric \HI\ radial velocity in
\kms\ taken at the midpoint of the \HI\ profile at the 20\% level.
The velocity is given in the optical convention $V = c \cdot
(\lambda-\lambda_o)/\lambda_o$. 

A ``:'' indicates an uncertain value whereas a ``*'' means
that a signal was measured off the center of the beam.

{\bf Col. 15:} Velocity width in \kms\ of the \HI\ profile measured
at the $50\%$ level of the peak intensity.

{\bf Col. 16:} Velocity width in \kms\ of the \HI\ profile measured
at the $20\%$ level of the peak intensity.

{\bf Col. 17:} \HI\ flux integral, in Jy\,\kms, uncorrected for
finite beam size.

{\bf Col. 18:} RMS noise level in m\,Jy measured over the region
used to fit a baseline, typically of a width of 1500~\kms\ centered
on, but not including, the detection.

\begin{landscape}  
\begin{table}[h]
 \normalsize
 \renewcommand{\baselinestretch}{0.65}
\caption{\HI-detections in the Hydra/Antlia region}
\label{det}
\scriptsize  
\hfill
\vfill
\begin{tabular*}{22.5cm}{
  l  @{\extracolsep{4mm}} p{1.4cm} @{\extracolsep{1mm}}
  c @{\extracolsep{2mm}}                                                                                                                                                   
  l@{\extracolsep{3mm}} l @{\extracolsep{4mm}}
  r @{\extracolsep{2mm}}r @{\extracolsep{4mm}} 
  p{2.7mm} @{\extracolsep{-1mm}}
  p{2.7mm} @{\extracolsep{-1mm}}
  p{2.7mm} @{\extracolsep{-1mm}} 
  p{2.7mm} @{\extracolsep{0mm}}
  p{6mm} @{\extracolsep{0mm}} p{4.5mm} @{\extracolsep{3mm}}
  r @{\extracolsep{2mm}} c @{\extracolsep{2mm}} 
  r @{\extracolsep{4mm}} r @{\extracolsep{4mm}} 
%
 r @{\extracolsep{0.5mm}} c @{\extracolsep{2mm}}
 r @{\extracolsep{0.5mm}} c @{\extracolsep{2mm}}
 r @{\extracolsep{0.5mm}} c @{\extracolsep{1mm}}
 r @{\extracolsep{0.5mm}} c @{\extracolsep{1mm}}
 r @{\extracolsep{2mm}}
 r @{\extracolsep{0mm}}
}
\hline
\vspace{-1mm} \\
 Ident. & Other & IR & \ \ \ \ R.A. & \ \ \ Dec.& gal $\ell$ \ & gal $b$ \ &
 \multicolumn{4}{l}{Type} & 
 \multicolumn{2}{c}{$D$ x $d$} & 
 $B_{J}$ & 
 $E_{(B-V)}$ & 
 $D_{J}^0$ &
 $B_{J}^0$ &
 {$V^{opt}_{hel}$} & &
 {$\Delta V_{50}$} & &
 {$\Delta V_{20}$} & &
 {$I \ \ $} & &
 {$rms$} &
 $V_{other}$ \\
& &  &
(h\,\, m\,\, s) & ($\deg$\,\, $\arcmin$\,\,$\arcsec$) & ($\deg$) \ &($\deg$) \ &
& & & &
\multicolumn{2}{c}{($\arcsec$)} & ($^{\rm m}$) & ($^{\rm m}$) &
($\arcsec$) & ($^{\rm m}$) & 
km/s & & km/s & & km/s & & {Jy\,km/s} & & m\,Jy & km/s \\
\vspace{-1mm} \\
\ \ \ (1) & \ \ \ (2) & (3) & \ \ \ \ (4) & \ \ \ \ (5) & (6) \ & (7) \
 & \multicolumn{4}{c}{(8)} \ & \multicolumn{2}{c}{(9)} \ & \ \ (10) &
 {(11)} & (12) & (13) & (14) & & (15) & & (16) & & (17) & & (18)  & (19) \ \ \\
\hline
\vspace{-1mm} \\
RKK0347 & FGCE0717 &    & 08 42 36.8 & -55 23 34 &  272.44 &  -8.03 & S& & & & \hfill  60x&\hfill   7 & 17.2 & 0.28 &  81 & 15.9 &  2786 & & 106& & 129& &  1.60& &  2.2&             \\
RKK0718 &          &    & 08 56 04.5 & -55 04 02 &  273.39 &  -6.32 & S& & & & \hfill  67x&\hfill  60 & 15.1 & 0.38 & 108 & 13.4 &  3905 & & 127& & 164& &  1.38& &  2.8&             \\
RKK0764 & L165-008 &    & 08 57 54.2 & -53 52 31 &  272.65 &  -5.35 & D& & & & \hfill  60x&\hfill  54 & 15.1 & 0.41 & 109 & 13.2 &  5802 & & 164& & 195& &  3.97& &  6.3&             \\
RKK1037\,O&          &    & 08 57 45.  & -61 49.0 \ :&278.78 & -10.47 &  & & & & \hfill     &\hfill     &      & 0.19 &     &      &  2881 &*& 133&*& 149&*&  3.23&*&  6.8&             \\
RKK1037 &          &    & 09 08 18.8 & -61 49 35 &  279.61 &  -9.54 &
S& & &4& \hfill  67x&\hfill  60 & 15.0 & 0.21 &  80 & 14.1 &  2242 & & 128& & 147& &  2.94& &  4.5&             \\
RKK1287 & L091-010 &    & 09 23 28.7 & -62 52 56 &  281.61 &  -8.98 & S& & &3& \hfill  74x&\hfill  20 & 15.5 & 0.26 &  97 & 14.3 &  4562& & 268& & 291& &  7.42& &  2.9&             \\
RKK1288$^a$& L091-011 &  I & 09 23 26.8 & -63 40 46 &  282.18 &  -9.54 & S& & &3& \hfill 101x&\hfill  12 & 15.8 & 0.22 & 124 & 14.8 &  3197&*& 277&*& 306&*& 13.04&*&   5.2&     3192 (1)\\
RKK1312 &          &    & 09 25 19.0 & -59 30 14 &  279.36 &  -6.43 & S& & & & \hfill  54x&\hfill   7 & 17.1 & 0.36 &  85 & 15.4 &  2973& & 208& & 244& &  5.35& &  5.4&             \\
RKK1328 & L126-011 &    & 09 26 24.3 & -60 36 54 &  280.25 &  -7.12 & S& & &L& \hfill  74x&\hfill  27 & 15.7 & 0.28 &  99 & 14.4 &  2119& & 124& & 143& & 21.63& &  7.6&             \\
RKK1371 & L126-014 &    & 09 28 26.9 & -60 48 05 &  280.56 &  -7.08 & E& & &5& \hfill  87x&\hfill  47 & 14.3 & 0.29 & 126 & 13.0 &  2196& & 263& & 290& &  8.33& &  6.9&     2273 (1)\\
\vspace{-1.20mm} \\                                                                                                                                                        
RKK1402 & L091-014 &    & 09 30 12.9 & -63 05 13 &  282.31 &  -8.59 & S& & &7& \hfill  74x&\hfill  27 & 15.6 & 0.26 &  95 & 14.5 &  3047& & 188& & 213& &  3.81& &  3.9&             \\
RKK1419 & L126-018 &    & 09 31 09.6 & -59 47 53 &  280.10 &  -6.13 & S& & &L& \hfill  40x&\hfill  16 & 16.8 & 0.44 &  75 & 14.7 &  2955& & 193& & 229& &  6.15& &  5.1&             \\
RKK1445 & FGCE0751 &    & 09 32 35.7 & -59 53 34 &  280.30 &  -6.07 & S& & & & \hfill  87x&\hfill   4 & 17.1 & 0.39 & 144 & 15.3 &  2978& & 189& & 207& &  5.98& &  4.5&             \\
RKK1464 & L126-019 &  I & 09 34 13.5 & -61 16 59 &  281.40 &  -6.95 & S& & &6& \hfill 108x&\hfill  81 & 14.2 & 0.36 & 168 & 12.6 &  2625& & 156& & 184& & 35.45& &  5.2&     2569 (2)\\
RKK1479 & L126-020 &    & 09 35 50.2 & -62 00 21 &  282.04 &  -7.35 & S& & &5& \hfill  54x&\hfill  47 & 15.3 & 0.38 &  90 & 13.5 &  2818& & 171& & 187& & 10.75& &  5.2&             \\
RKK1486 & L091-015 &    & 09 36 32.4 & -63 56 44 &  283.42 &  -8.73 & S& & &7& \hfill  81x&\hfill  40 & 15.1 & 0.25 & 103 & 14.0 &  2932& & 233& & 254& &  7.86& &  3.3&             \\
RKK1492 & L126-022 &    & 09 37 00.9 & -60 55 55 &  281.41 &  -6.47 & S&B& &4& \hfill  81x&\hfill  34 & 14.9 & 0.38 & 134 & 13.1 &  2787& & 268& & 291& & 11.26& &  4.0&     2918 (3)\\
RKK1499 &          &    & 09 37 30.7 & -60 18 49 &  281.04 &  -5.97 & S& & &L& \hfill  54x&\hfill   5 & 17.6 & 0.34 &  80 & 16.1 &  3077& & 160& & 177& &  4.90& &  5.2&             \\
RKK1541 &          &    & 09 40 03.7 & -64 13 32 &  283.90 &  -8.68 & S& & &M& \hfill  54x&\hfill   8 & 17.2 & 0.28 &  72 & 16.0 &  4589& & 372& & 415& &  4.44& &  3.8&             \\
RKK1561 &          &  I & 09 41 44.0 & -60 47 54 &  281.75 &  -5.98 & E& & &4& \hfill  34x&\hfill  20 & 16.2 & 0.46 &  67 & 14.1 &  3107& & 270& & 299& &  9.38& &  4.9&     3070 (1)\\
\vspace{-1.20mm} \\                                                                                                                                                             
RKK1610 & HIZSS056 &    & 09 45 24.8 & -48 08 29 &  273.90 &   3.96 &  & & & & \hfill  54x&\hfill  13 & 16.9 & 0.48 & 117 & 14.5 &   878& &  75& &  94& &  8.25& & 11.8&      880 (7)\\
RKK1717 &          &    & 09 51 43.2 & -63 14 36 &  284.24 &  -7.09 &  & & & & \hfill  34x&\hfill  30 & 16.4 & 0.26 &  44 & 15.2 &  3126& & 108& & 128& &  1.94& &  3.7&             \\
RKK1766 &          &    & 09 55 09.  & -60 40.3 \ :&282.94 &  -4.84 & S& & &4& \hfill  40x&\hfill  27 & 16.1 & 0.53 &  89 & 13.6 &  3294& & 252& & 265& &  9.45& &  5.9&             \\
RKK1856 & L092-003 &    & 09 59 28.3 & -67 18 18 &  287.45 &  -9.76 & S& & &1& \hfill 101x&\hfill  34 & 14.9 & 0.21 & 122 & 14.0 &  1491& & 182& & 200& &  6.19& &  7.5&     1478 (1)\\
RKK1862 & L092-004 &    & 09 59 45.1 & -67 38 33 &  287.68 & -10.01 & S& & &1& \hfill 114x&\hfill  47 & 14.6 & 0.24 & 143 & 13.5 &  1857& & 175& & 198& & 12.35& &  5.9&             \\
RKK1864 &          &    & 10 00 03.1 & -64 26 26 &  285.72 &  -7.46 & S& & &M& \hfill  81x&\hfill   8 & 17.0 & 0.24 & 102 & 15.9 &  9857& & 404& & 425& &  6.26& &  5.2&             \\
RKK1877 &          &    & 10 00 50.9 & -65 06 08 &  286.20 &  -7.93 &  & & & & \hfill  38x&\hfill  30 & 16.4 & 0.27 &  50 & 15.2 &  6040&:& 225&:& 251&:&  3.60&:&  5.2&             \\
RKK1909 & L092-007 &    & 10 03 32.6 & -67 26 54 &  287.85 &  -9.63 & I& & & & \hfill 134x&\hfill  34 & 15.0 & 0.21 & 161 & 14.1 &  2006& & 109& & 138& &  6.68& &  7.1&             \\
RKK1919 & AM1002   &    & 10 04 47.3 & -48 44 26 &  276.80 &   5.49 & S& & &6& \hfill  40x&\hfill  40 & 15.3 & 0.47 &  84 & 12.9 &  6714& & 198& & 210& &  2.13& &  3.4&             \\
RKK1919\,B& INV1     &    &            &           &         &        &  & & & & \hfill     &\hfill     &      &      &     &      & 10742& & 265& & 276& &  2.35& &  3.7&             \\
\vspace{-1.20mm} \\                                                                                                                                                             
RKK1938 & L062-002 &  I & 10 05 14.7 & -67 44 52 &  288.17 &  -9.78 & S&X& &3& \hfill  60x&\hfill  54 & 15.1 & 0.21 &  72 & 14.2 &  5196&:&  41&:&  75&:&  3.42&:&  7.7&     5194 (3)\\
RKK1939 & L213-006 &    & 10 06 06.6 & -50 36 38 &  278.09 &   4.11 & I& & & & \hfill  54x&\hfill  16 & 16.0 & 0.44 & 106 & 13.8 &  5299& & 296& & 317& &  8.01& &  4.6&     5262 (1)\\
RKK1945 &          &    & 10 06 29.6 & -50 14 33 &  277.92 &   4.44 & S&B&R&4& \hfill  47x&\hfill  40 & 15.1 & 0.44 &  93 & 12.9 &  5268& & 246& & 283& &  6.37& &  5.4&             \\
RKK1947 & L092-009 &    & 10 06 02.1 & -67 03 48 &  287.81 &  -9.18 & I& & & & \hfill  81x&\hfill  13 & 16.3 & 0.24 & 101 & 15.2 &  1461& & 123& & 144& &  4.82& &  5.2&     1455 (9)\\
        & RKK1959  &    & 10 06 37.3 & -67 06 49 &  287.89 &  -9.19 &  & & & & \hfill  20x&\hfill  20 & 17.6 & 0.23 &  25 & 16.6 &  1623&*&  75&*&  92&*&  2.32&*&  5.2&     1612 (9)\\
        & Obj.B    &    & 10 06 20.  & -67 09.2 \ :&287.89 &  -9.24 &  & & & & \hfill     &\hfill     &      & 0.23 &     &      &  1743&*&  46&*&  78&*&  2.38&*&  5.2&     1721 (9)\\
        & INV 2    &    & 10 04 45.  & -66 56.0 \ :&287.63 &  -9.15 &  & & & & \hfill     &\hfill     &      & 0.22 &     &      &  1877&*&  65&*& 155&*&  2.61&*&  5.2&             \\
RKK2043 & L092-015 &  I & 10 11 00.6 & -67 08 40 &  288.26 &  -8.96 & S&X& &3& \hfill  60x&\hfill  47 & 15.0 & 0.19 &  70 & 14.2 &  5252& & 191& & 211& &  3.84& &  5.4&     5169 (2)\\
RKK2063 & L127-011 &  I & 10 12 11.9 & -62 32 03 &  285.69 &  -5.12 & S& & &3& \hfill  67x&\hfill  27 & 15.1 & 0.29 &  92 & 13.8 &  3403& & 238& & 376& &  4.19& &  5.9&     3370 (4)\\
RKK2097 &          &    & 10 13 10.1 & -62 16 52 &  285.63 &  -4.85 &  & & & & \hfill  40x&\hfill  24 & 15.6 & 0.35 &  62 & 14.0 &  3784&:& 188&:& 235&:&  1.62&:&  4.4&             \\
\vspace{-1.20mm} \\                                                                                                                                                             
RKK2136 &          &    & 10 14 08.8 & -65 30 13 &  287.57 &  -7.44 & S& & &M& \hfill  74x&\hfill  22 & 16.0 & 0.28 & 100 & 14.7 &  5858& & 261& & 275& &  7.74& &  7.8&             \\
RKK2192 & L092-019 &  I & 10 16 16.3 & -64 51 60 &  287.39 &  -6.79 & S& & & & \hfill  67x&\hfill   8 & 16.9 & 0.29 &  92 & 15.6 &  3884& & 228& & 241& &  5.40& &  6.5&             \\
RKK2247 &          &    & 10 17 47.0 & -65 52 03 &  288.08 &  -7.53 & S&X& &M& \hfill  60x&\hfill  40 & 16.0 & 0.29 &  82 & 14.7 &  4588& & 123& & 169& &  3.21& &  6.1&             \\
RKK2249 &          &    & 10 17 59.0 & -62 45 07 &  286.36 &  -4.93 & S& & &5& \hfill  60x&\hfill   7 & 17.3 & 0.39 & 103 & 15.4 &  3421& & 201& & 218& &  4.77& &  7.4&             \\
RKK2276 &          &    & 10 18 33.5 & -63 37 25 &  286.90 &  -5.62 & S& & &L& \hfill  60x&\hfill   8 & 17.3 & 0.36 &  95 & 15.6 &  3615& & 307& & 338& &  3.73& &  5.3&             \\
RKK2426 & L214-002 &  I & 10 23 11.6 & -49 28 14 &  279.71 &   6.60 & S& & &2& \hfill  77x&\hfill  44 & 14.8 & 0.30 & 105 & 13.5 &  5443& & 335& & 364& & 17.58& & 11.9&     5411 (5)\\
RKK2507 &          &  I & 10 25 58.8 & -53 52 21 &  282.44 &   3.11 & S&B& &5& \hfill  51x&\hfill  34 & 15.4 & 0.55 & 133 & 12.5 &  6805& &  93& & 115& &  4.67& &  8.7&             \\
RKK2508 & L062-010 &  I & 10 25 33.2 & -67 44 43 &  289.77 &  -8.68 & S&X& &7& \hfill  60x&\hfill  60 & 14.9 & 0.25 &  76 & 13.8 &  6030&:&  88&:& 100&:&  5.18&:&  9.3&     6011 (1)\\
RKK2546$^a$&          &    & 10 27 19.8 & -49 06 57 &  280.10 &   7.26 & S& & &2& \hfill  67x&\hfill  16 & 16.1 & 0.31 &  93 & 14.7 &  5584&*& 172&*& 187&*&  2.25&*&  7.6&             \\
RKK2585& L214-008 &  I & 10 28 25.1 & -50 41 29 &  281.08 &   6.01 & S& & &M& \hfill  65x&\hfill  47 & 14.9 & 0.29 &  90 & 13.6 &  7005& & 373& & 391& &  5.59& &  6.2&     6868 (6)\\
\vspace{-1.20mm} \\                                                                                                                                                             
RKK2652 & L214-013 &  I & 10 31 01.4 & -49 02 51 &  280.58 &   7.63 & S& & &3& \hfill  87x&\hfill  48 & 14.5 & 0.33 & 129 & 13.0 &  4886& & 324& & 353& &  5.29& &  4.1&     4780 (5)\\
RKK2684& L092-022 &  I & 10 31 57.4 & -63 42 30 &  288.21 &  -4.90 & S& & &7& \hfill 128x&\hfill  47 & 14.6 & 0.48 & 266 & 12.2 &  3762& & 318& & 341& & 14.05& &  7.1&     3671 (1)\\
RKK2687 &          &    & 10 32 20.0 & -50 42 30 &  281.62 &   6.32 & S& & &M& \hfill  51x&\hfill  38 & 15.8 & 0.30 &  72 & 14.4 &  6097& & 138& & 155& &  2.01& &  5.0&             \\
RKK2740 &          &  I & 10 35 52.5 & -54 06 28 &  283.81 &   3.65 & S& & &5& \hfill  54x&\hfill   5 & 17.6 & 0.52 & 124 & 15.0 &  6285&:& 359&:& 436&:&  6.78&:&  6.2&             \\
RKK2754 & L168-009 &  I & 10 37 18.0 & -54 56 07 &  284.40 &   3.04 & S& & &5& \hfill  54x&\hfill  24 & 16.1 & 0.73 & 273 & 11.7 &  2674& & 306& & 327& & 19.47& &  5.6&     2754 (7)\\
RKK2763 & L214-016 &    & 10 38 09.5 & -50 09 33 &  282.15 &   7.26 & S& & &4& \hfill 148x&\hfill  15 & 15.8 & 0.53 & 359 & 13.1 &  7074& & 436& & 469& &  8.67& &  4.5&             \\
RKK2771 &          &    & 10 38 09.9 & -66 06 13 &  289.99 &  -6.64 & S& & & & \hfill  47x&\hfill  43 & 16.2 & 0.38 &  77 & 14.4 &  7558& & 209& & 241& &  7.76& &  6.2&             \\
RKK2847 & L169-002 &  I & 10 48 14.9 & -53 18 10 &  285.04 &   5.23 & S& & &L& \hfill  94x&\hfill  27 & 15.2 & 0.50 & 204 & 12.7 &  2683& & 301& & 324& & 10.26& &  6.2&     2712 (1)\\
RKK2879 & L169-005 &    & 11 02 31.3 & -53 38 36 &  287.11 &   5.85 & S&B& &1& \hfill  67x&\hfill  47 & 14.8 & 0.34 &  96 & 13.4 &  3917& & 386& & 408& &  4.66& &  5.5&     3766 (1)\\
RKK2991 & L170-002 &  I & 11 26 03.7 & -52 46 47 &  290.09 &   7.95 & S&B&R&5& \hfill  81x&\hfill  81 & 14.5 & 0.26 & 106 & 13.3 &  5192& & 204& & 232& &  8.07& &  3.9&     4941 (8)\\
\vspace{-1.20mm} \\                                                                                                                                                    
RKK2992 & L170-003 &    & 11 26 08.6 & -54 14 08 &  290.58 &   6.58 & S& & &2& \hfill  94x&\hfill  40 & 14.6 & 0.26 & 125 & 13.4 &  5697& & 480& & 516& & 16.90& &  5.7&     5369 (1)\\
RKK3017 &          &    & 11 28 59.7 & -58 25 21 &  292.30 &   2.74 & S&X& &3& \hfill  74x&\hfill  34 & 15.3 & 0.63 & 255 & 11.8 &  4531& & 516& & 542& &  4.88& &  4.7&     4639 (1)\\
RKK3165 & L170-009 &  I & 11 39 16.4 & -53 33 12 &  292.23 &   7.81 & S& & &2& \hfill  74x&\hfill  27 & 15.8 & 0.22 &  90 & 14.8 &  5027& & 347& & 369& &  4.45& &  5.2&             \\
RKK3176 & L170-010 &  I & 11 40 04.7 & -54 23 51 &  292.58 &   7.03 & S&B& &4& \hfill  60x&\hfill  54 & 15.1 & 0.29 &  82 & 13.8 &  4558& & 273& & 303& &  7.75& &  3.6&     4677 (8)\\
RKK3204 & L216-035 &    & 11 44 04.3 & -52 36 39 &  292.67 &   8.91 & S& &S&5& \hfill  74x&\hfill  67 & 14.9 & 0.20 &  88 & 14.0 &  4538& & 153& & 178& &  3.98& &  3.9&     4446 (6)\\
RKK3245 & L170-012 &    & 11 47 57.8 & -53 54 13 &  293.57 &   7.81 & S& & &5& \hfill  47x&\hfill  34 & 16.0 & 0.22 &  58 & 15.0 &  5067& & 243& & 277& &  7.63& &  5.2&             \\
\vspace{-1.  mm} \\
\hline
\end{tabular*}
 \normalsize
\label{hidet}
\end{table}
\end{landscape} 

\begin{figure*}[p] 
\epsfig{file=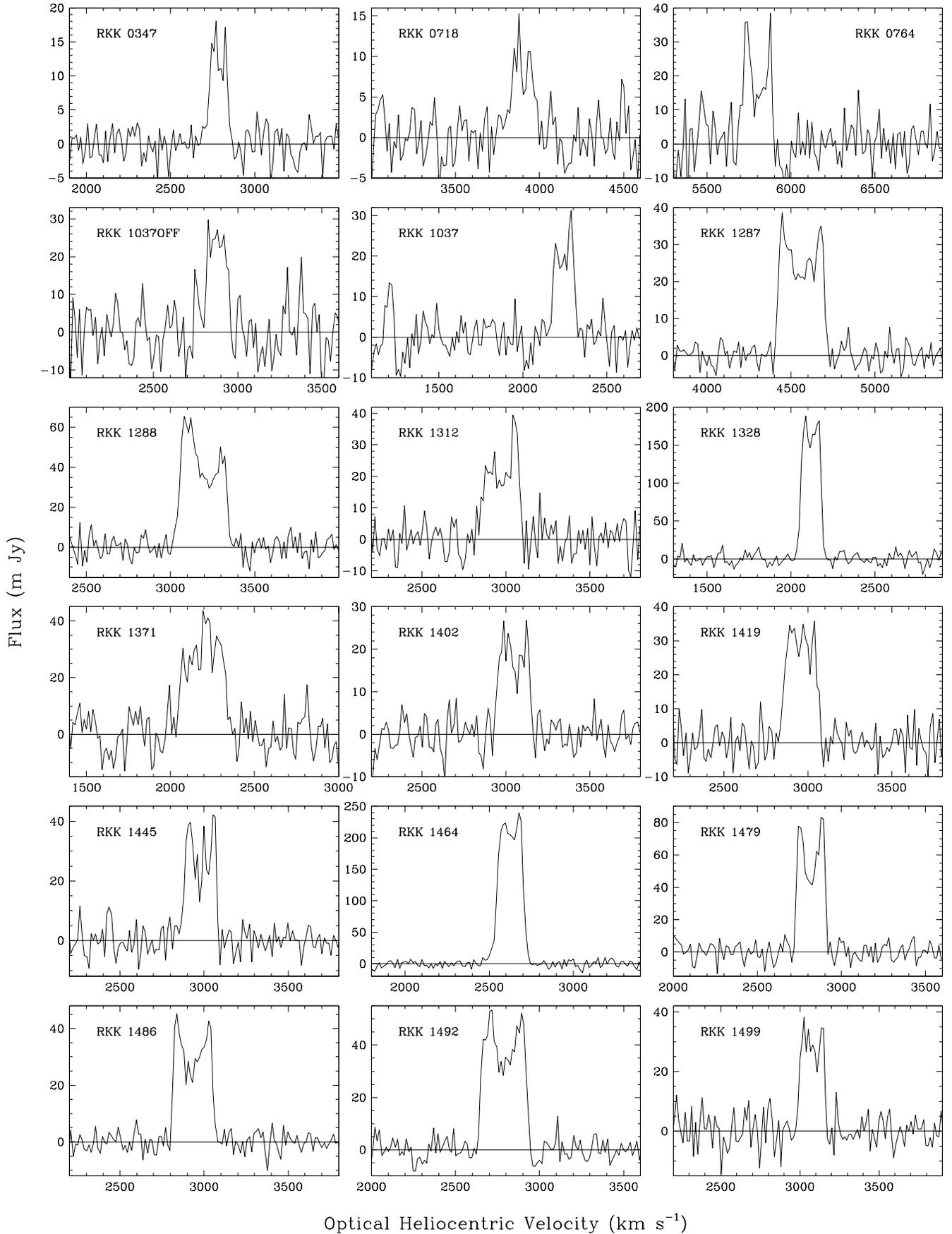,width=17.8cm,bbllx=41pt,bblly=63pt,
bburx=564pt,bbury=742,clip=,angle=0} 
\\[-0.5cm]
\caption{\HI\ profiles of the 66 \HI\ detections. The vertical axis
gives the flux density in m\,Jy, the horizontal axis the velocity
range (optical convention), generally centered on the radio velocity of
the galaxy displaying a width of 1600\kms.  All spectra are
baseline-subtracted and generally Hanning-smoothed once.  
The respective identifications are given within the panels.
}
\label{profile}
\end{figure*}
\addtocounter{figure}{-1} 
\begin{figure*}
\epsfig{file=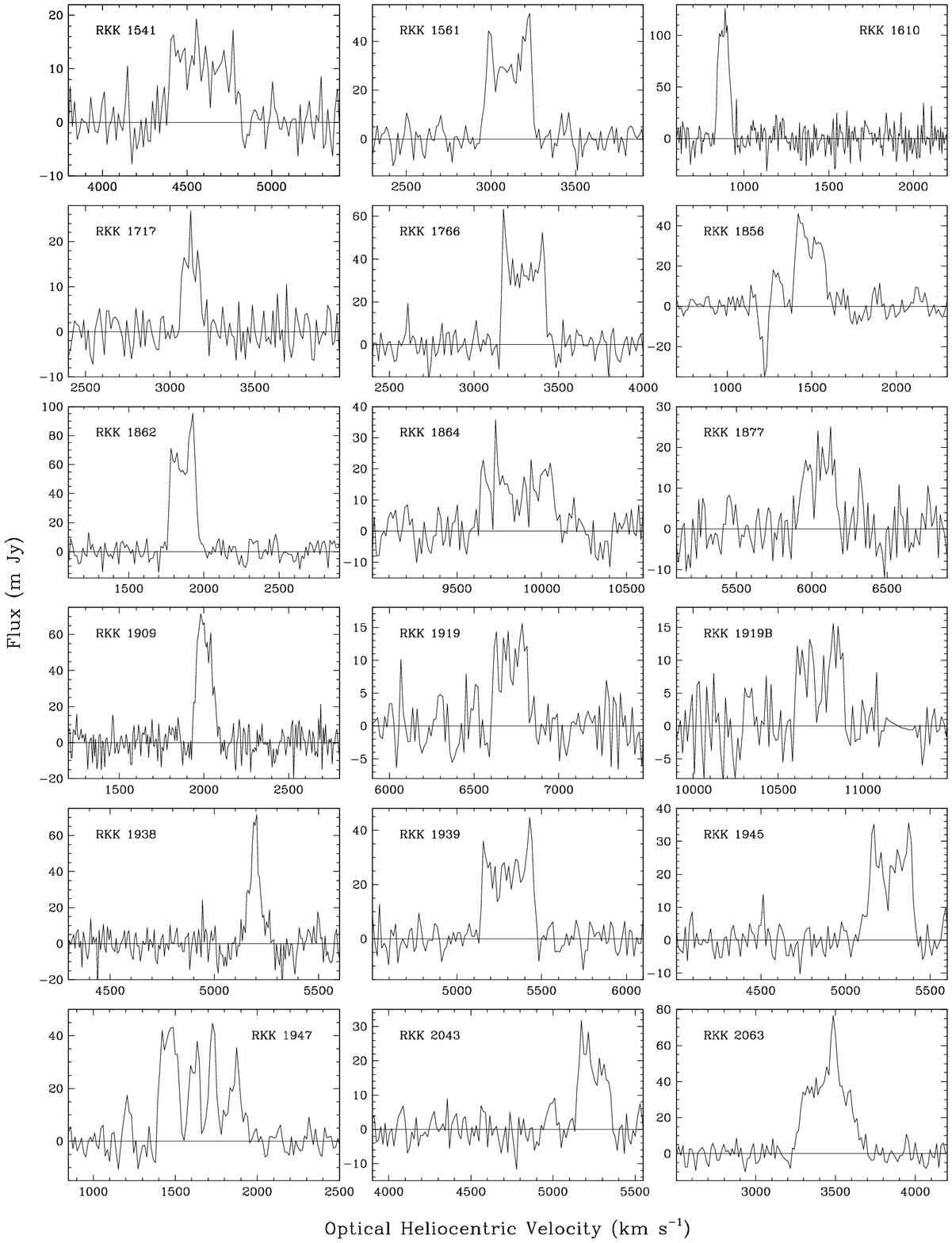,width=17.8cm,bbllx=41pt,bblly=63pt,
bburx=564pt,bbury=742,clip=,angle=0}
    \caption{ -- cont. Note the an interference feature present 
in the spectrum of RKK\,1919B has been excised.}
\end{figure*}
\addtocounter{figure}{-1} 
\begin{figure*}
\epsfig{file=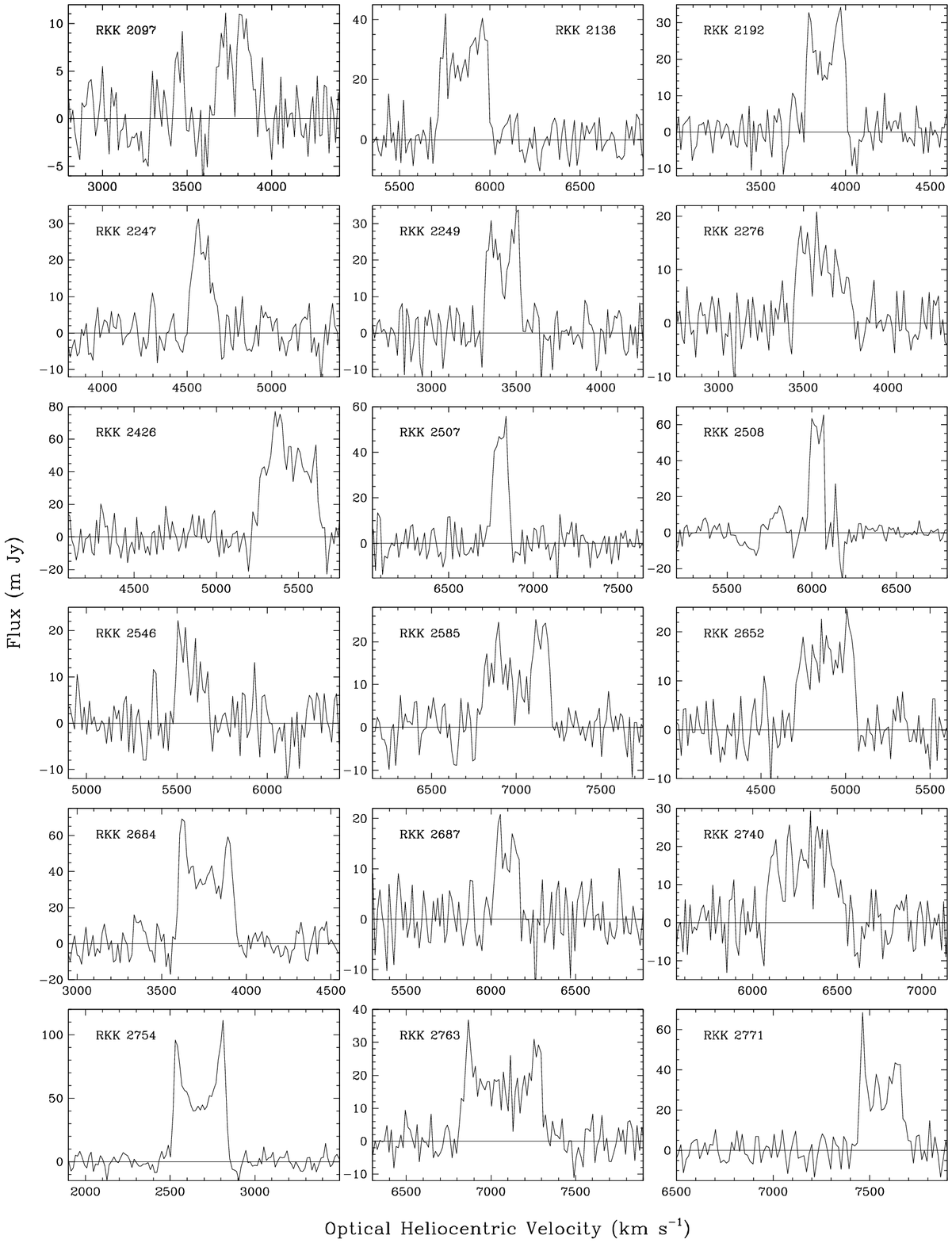,width=17.8cm,bbllx=41pt,bblly=63pt,
bburx=564pt,bbury=742,clip=,angle=0}
    \caption{ -- cont.}
\end{figure*}
\addtocounter{figure}{-1} 
\begin{figure*}
\epsfig{file=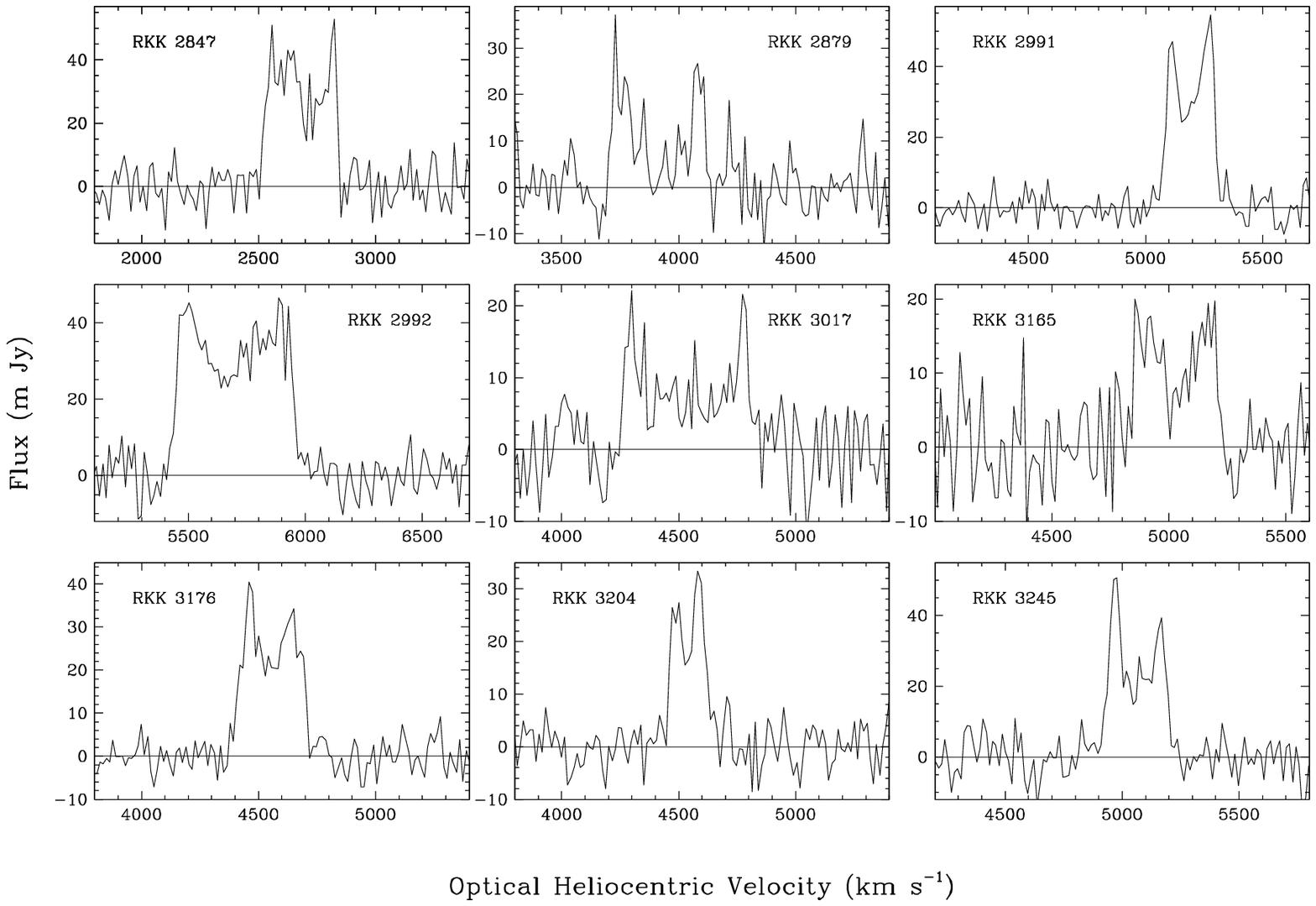,width=17.8cm,bbllx=41pt,bblly=385pt,
bburx=564pt,bbury=742,clip=,angle=0}
    \caption{ -- cont.}
\end{figure*}

{\bf Col. 19:} Independent velocity determinations come from the
following sources:

(1) Kraan-Korteweg et al. (1995) 

(2) Visvanathan \& Yamada (1996) 

(3) Fairall et al. (1998) 

(4) Strauss et al. (1992) 

(5) Fisher et al. (1995)

(6) The NED data compilation (prior to 1992)

(7) Henning et al. (2000)

(8) Dressler (1991)

(9) Gordon (in prep.), Gordon \etal\ (in prep.)\\
Obviously we did not repeat observations at
Parkes for galaxies with existing \HI\ data. Therefore, most of
the independent velocities listed in column 19 originate from optical
spectroscopy. Overall, the agreement with the radio data is excellent.

The two strong \HI\ sources RKK1610 and RKK2754 were also detected
with the shallow ZOA \HI\ survey (HIZSS\,056 and HIZSS\,064). The \HI\
parameters measured by the HIZSS are in very good agreement with our
pointed observations.

\subsection{Special cases} \label{weirdt}

\subsubsection{An OFF-detection in the RKK\,1037 beam}

The detection labelled RKK\,1037OFF in Fig.~\ref{profile} was first
seen as an unexpected negative signal in the scan directed towards
RKK\,1037.  A pointed observation towards the reference (or ``OFF'')
position at a 10\fm5 lower RA then yielded the equivalent positive
signal which is displayed in Fig.~\ref{profile}. 

At this position no galaxy is known in the literature. With its
Galactic latitude of $b=-10\fdg47$ it lies outside the optically
searched ZOA region. Considering that the optical extinction at this
position, $A_{\rm B} = 0\fm79$, is quite low, we are confident that such a
spiral galaxy should be visible.  Indeed, its \HI\ mass of $>$\,$1.6
\cdot 10^9\msun$ (the inequality is due to the galaxy's not lying at
the center of the beam, where sensitivity is at a maximum) would
suggest for a typical spiral galaxy an absolute magnitude of at least
as bright as $M_{\rm B} \la -18\fm5$.  Using its velocity $V_o = 2588$\kms\
corrected to the centroid of the Local Group following Yahil, Sandage
\& Tammann (1977) and a Hubble constant of $H_{rm 0} = 50$\kms\,Mpc$^{-1}$
gives a value of $B^{\rm o} \la 15\fm1$ and, when observed through the
extinction layer at that position, $B \la 15\fm9$.  

A close inspection of the ESO/SERC IIIaJ film copy of the field F125
revealed four galaxies (C1 $-$ C4) in the vicinity of the OFF-pointing.
Their likelihood of being the optical counterpart for the \HI\ emission
will be discussed in order of increasing distance from
the pointing.

{\bf C1:} Closest to the center of the beam, at only $1\farcm7$ from
the OFF pointing, we found the first galaxy at RA $=08^{\rm h}57^{\rm
m}48^{\rm s}$, Dec$ = -61\deg 47\arcmin 16\arcsec$ (J2000.0). However,
this galaxy seems an unlikely counterpart. It is a very small,
edge-on spiral of very low-surface brightness. With an estimated
magnitude of $B_{\rm J} \sim 19\fm5$, this object does not match 
the values expected for this spiral galaxy at all. Furthermore
the profile is atypical for an edge-on galaxy.

{\bf C2:} At a distance of $7\farcm6$ another faint very low surface
brightness, edge-on spiral galaxy was identified (C2 at $08^{\rm
h}58^{\rm m}45^{\rm s}$, $-61\deg 46\arcmin 18\arcsec$). As it is has
a similar magnitude and size as C1, it is equally improbable to have
produced the observed \HI\ emission, particularly as the observed flux
is underestimated by a factor of 2 for the offset of $7\farcm6$ from the
pointing center.

{\bf C3:} Close to C2, at 7\farcm7 from the pointing, we found another
possible galaxy counterpart at $08^{\rm h}58^{\rm m}49.9^{\rm s}$,
$-61\deg 48\arcmin 03\arcsec$. With a size of about $20\arcsec \times
12\arcsec$, we estimate its magnitude to be of the order of $B_{\rm J} \sim
18\fm0$. However, this galaxy looks more like an early type galaxy,
possibly S0 or S0/a. This is independently confirmed by near infrared
DENIS\footnote{The DEep Near-Infrared Survey of the Southern Sky
(Epchtein 1997, Epchtein \etal 1997, Fouqu\'e et al. 2000)} images
which were available to us and which we have also consulted for
possible cross-identifications. It is the only counterpart seen on the
NIR DENIS survey, therewith confirming its early-type morphology
through its intrinsic red color. We derived an $I$-band magnitude for
C3 with the automated galaxy pipeline for DENIS (Mamon \etal\ 1997) of
$I = 16\fm41 \pm 0\fm12$. Taking account of extinction effects and
intrinsic color, the $I$-band magnitude and $B_{\rm J}$ estimate agree
reasonably well. However, with its probable early-type morphology and
magnitude, C3 does not match the expectations from the observed \HI\ flux
either even when corrected for the positional offset ($f=2$).

{\bf C4:} A final candidate was found at a considerable distance
($\Delta r = 11\farcm5$) from the beam center. At RA\,$=08^{\rm
h}59^{\rm m}21.2^{\rm s}$, Dec\,$ =-61\deg 47\arcmin 38\arcsec$, this
object is a prominent (at least by ZOA standards) nearly edge-on
spiral galaxy.  It measures $35\arcsec \times 15\arcsec$. With an
intermediate surface-brightness, we estimate its magnitude in the blue
as $B_{\rm J} \sim 17\fm0$, respectively $B_{\rm J}^{\rm o} \sim 16\fm2$. This is close
to the expected value deduced from the \HI\ observations, even after
the flux is corrected with the factor 5 due to the offset of
$11\farcm5$ from the center of the beam. With a morphology of $\sim$
Sbc, this galaxy thus seems the most likely counterpart of the
RKK\,1037OFF detection.

\subsubsection{Two detections in the RKK\,1919 beam}

The pointing towards RKK\,1919 revealed two well-separated 21-cm
detections, one at $V_{hel} = 6567$\kms\ and one at 10\,370\kms, both
consistently recovered in the individual scans. No other galaxy
besides the one described in Table~\ref{det} is visible on the sky
surveys within the area of the HPBW of the Parkes antenna, and no
independent redshift measurement is available to assign one of the two
values to the optical galaxy.  Determining and comparing the absolute
parameters for the two redshift values make the nearer redshift the
more likely value for the optically visible galaxy. Furthermore, a
low-surface brightness galaxy at a redshift of over 10\,000\kms\
behind an obscuration layer of $A_{\rm B} = 1\fm95$ might easily remain
invisible in the optical.

\subsubsection{Four detections within the beam of ESO\,92-9}

\begin{figure}
\psfig{file=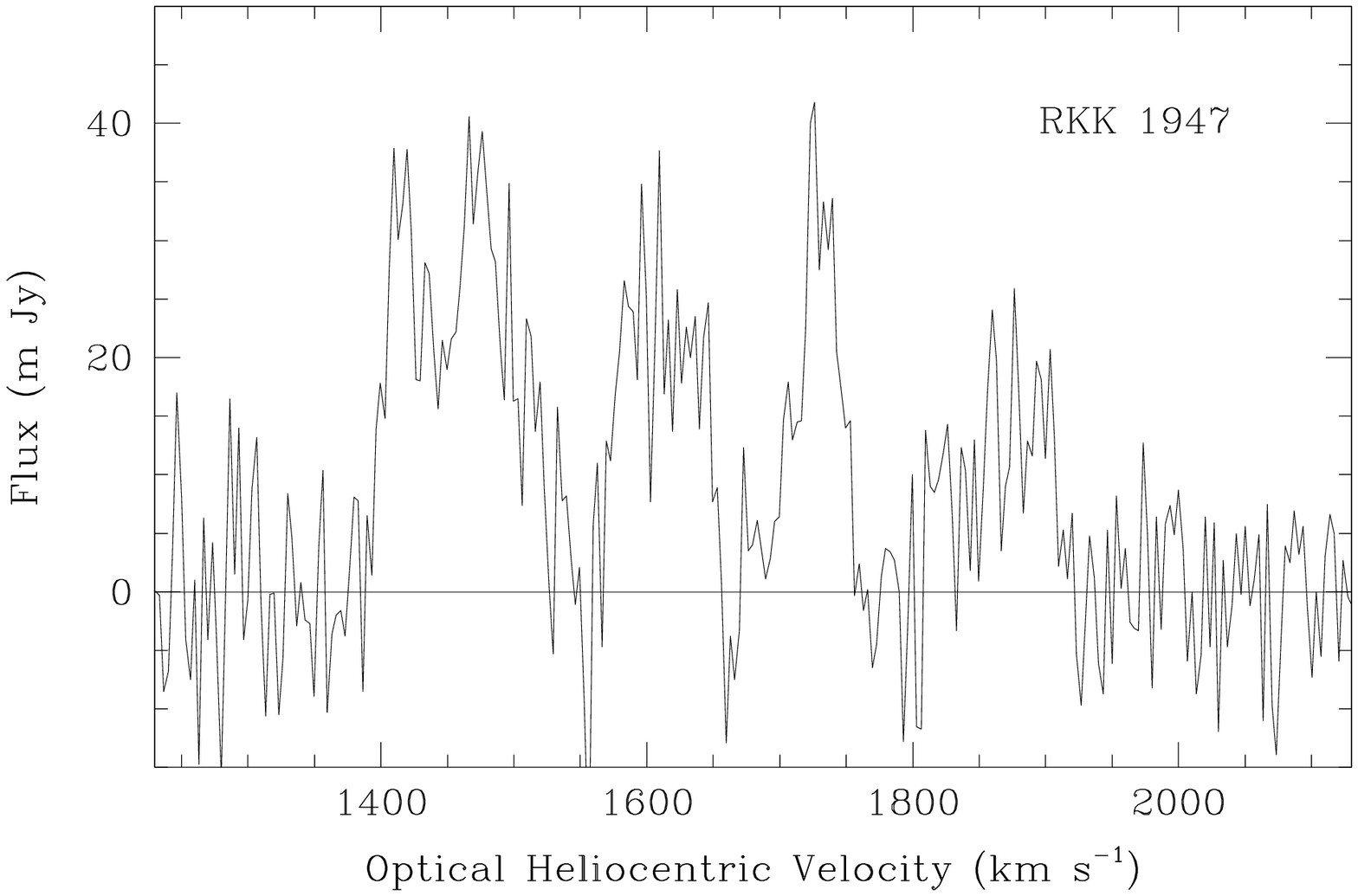,width=8.8cm}
    \caption{\HI\ spectrum of the RKK\,1947 region reobserved with higher
    resolution: the 2048 channels distributed over 16\,MHz lead to a channel 
    spacing of 1.7\kms, or 3.4\kms\ after Hanning smoothing as
    shown here. All four previously identified peaks are confirmed as
    individual discrete signals.
}
\label{weird}
\end{figure}

The spectrum obtained at the pointing towards RKK\,1947 = ESO\,92-9
reveals four distinct peaks in the velocity range of $1400 \la V_{\rm hel}
\la 1900$\kms, one peak adjacent to the next one and not, or just
barely, overlapping (see Figs.~\ref{profile} and \ref{weird}). In order
to confirm the reality and the discreteness of the four individual
peaks, we reobserved this galaxy with higher resolution and a narrower
bandwidth (2048 channels over 16\,MHz centered on $V = 2000$\kms). The
resulting spectrum, which is shown in Fig.~\ref{weird}, clearly
confirms the four individual peaks -- although the highest redshift
one seems slightly weaker here -- and also suggests that the four
signatures seem to originate from four different sources.

\begin{figure}
\psfig{file=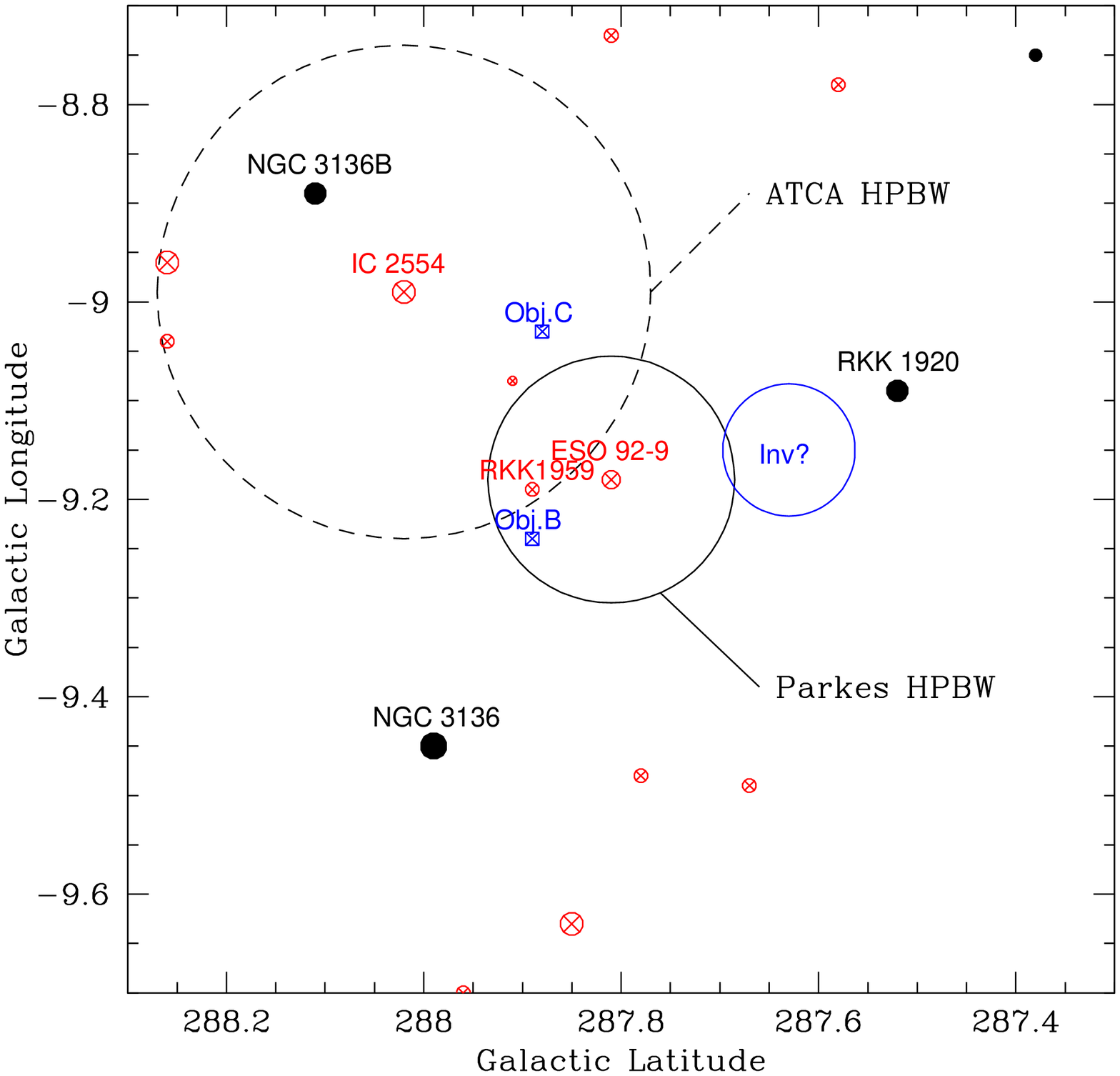,width=8.8cm} \caption{Distribution of
    galaxies within one square degree centered on RKK\,1947=ESO\,92-9
    with the solid circle marking the Parkes beam. The filled circles
    represent early type galaxies, the circled crosses spiral
    galaxies. The size of the symbols reflect the galaxies' apparent
    magnitudes. The squares are objects without optical counterparts
    discovered by ATCA observations of IC\,2554 (the dashed circle
    indicates the ATCA field) and the circle around ``Inv'' the
    location of the fourth peak in the HI profile (see Fig.~3), for
    which no optical counterpart has been found either.}
\label{weird_env}
\end{figure}

The region around ESO\,92-9 is quite a busy region with a number of
known nearby galaxies at similar velocities. However, only two spiral
galaxies, ESO\,92-9 itself, an edge-on spiral of $81\arcsec \times
13\arcsec$, and RKK\,1959, a small, face-on irregular galaxy
($20\arcsec \times 20\arcsec$), lie within or close to the beam of the
Parkes telescope (see solid circle centered on ESO\,92-9
in Fig.~\ref{weird_env}). These galaxies hence might be the sources of
two of the four peaks in the observed profile. No independent velocity
is known for either of these two spiral galaxies so it remains unclear
which \HI\ signal belongs to which galaxy.

A further interesting question was whether the remaining part of the
signal could be ascribed to the nearby ($\Delta r = 17\arcmin$) strong
\HI-emitting, interacting galaxy IC\,2554 with a flux of $I =
29.3$\,Jy\kms, a heliocentric velocity of $V_{\rm hel} = 1384$\kms, and
the broad linewidth of $\Delta V_{\rm 20} = 400$\kms\ (Huchtmeier \&
Richter 1989).

A clarification of this confusing area was possible only through the
ATCA observations of the interacting galaxy IC\,2554 (see dashed circle in
Fig.~\ref{weird_env}) by Gordon (in prep.), and Gordon \etal\ (in prep.) which
they kindly made available to us prior to publication. Their
observations revealed that none of the peaks in our spectrum are due
to IC\,2554. But the first two peaks centered on $V_{\rm hel} = 1405$\kms\
and 1605\kms\ are matched exactly by their detections (profiles and 
intensities) at the positions of ESO\,92-9 and RKK\,1959.

In their 21-cm maps they had two additional detections, although for two
of them they did not find an optical counterpart -- a bit surprising
for an extinction of less than 1 magnitude in the optical. But were they
responsible for the two higher velocity peaks? In fact, the data for their
Object B (Obj. B) resembled the third signal in our profile, however,
their Obj. C was at a distinctly higher velocity than our fourth peak.

Due to the fact that the fourth peak in the spectrum of
Figs.~\ref{profile} and \ref{weird} could be reproduced consistently
with our scans and that Gordon \etal\ did find two \HI\ clouds without
optical counterparts, we considered the option that maybe yet another
invisible \HI\ cloud could exist that gives rise to the last peak, an
\HI\ cloud beyond the ATCA beam but not too far away from the Parkes
beam.  Using the HIPASS (Staveley-Smith, Engel \& Webster 2000) we
inspected spectra at larger and larger distances from the two beams in
the elongation of the axis defined by IC\,2554 to ESO\,92-9, and
indeed could recover the same signal in the HIPASS data.  It is
strongest ($\sim 0.04$\,Jy) at RA $=10^{\rm h}04^{\rm m}45^{\rm s}$,
Dec$ = -66\deg 56\arcmin 00\arcsec$, respectively $\ell= 287\fdg63, b=
-9\fdg15$, with a positional uncertainty of $4\arcmin$.  Hence, we
believe it to be an \HI\ cloud in the direction of RKK\,1920, a
lenticular galaxy at 2081\kms. At an offset of $11\arcmin$ from the
center of the beam, the flux of this \HI\ cloud would be a factor of 5
higher compared to what we observed.

A renewed inspection of this whole area on the sky survey, as well as
the fits-file of the second generation red DSS image -- which permits
the detection of very low surface brightness features by varying the
contrast -- did unveil a previously unrecognized, extremely
low-surface brightness spiral galaxy of about $15\arcsec \times
5\arcsec$ and $B_{\rm J} \sim 19\fm5$ partly covered by foreground stars,
hence difficult to recognize, at RA $=10^{\rm h}07^{\rm m}25^{\rm s}$,
Dec$ = -67\deg 02\arcmin 52\arcsec$, ($\ell = 287\fdg91,b =
-9\fdg08$). This faint galaxy is indicated in Fig.~\ref{weird_env} as
the tiny circled cross without further name which lies just outside the
Parkes beam about $4\arcmin$ from Obj. C in the direction of IC\,2554.
It does not seem to be the counterpart for any of the three
\HI\ signals for which so far no luminous counterpart has been found,
i.e.: Obj. B and C by the ATCA observations of Gordon \etal, and
Obj. B and the new object labelled ``Inv'' evident in the Parkes
observations.

So we seem to have three isolated, optically invisible \HI\ clouds in this
highly interactive region - and possibly even more? This certainly is
a region that merits further sensitive \HI\ synthesis mapping, as
well as deep infrared imaging.

\section{The non-detections}

In Table~\ref{ndet} the galaxies that were observed in the
Hydra/Antlia region but not detected (N = 82) are listed with the
searched velocity range as well as the rms within that interval.  Some
spectra did reveal a signal, but careful investigation showed them to
be due to close neighbors rather than the targeted galaxies. These few
cases are marked with the superscript ``$^{\rm a}$'' in Column 1 of
Table~\ref{ndet} and discussed in further detail in
Sect.~\ref{ndetsig}.

{\bf Col. 1 - 13:} Same explanations as given for Table~\ref{det},
where the superscript ``$^{\rm a}$'' in column 1 indicates that a signal of
another galaxy is visible in the scan (see also Table~\ref{det}).

{\bf Col. 14:} The searched velocity range.

{\bf Col. 15:} The rms noise of the searched velocity range,
typically \ of \ the order of \ 3 m\,Jy. These \ values were determined \,
after \, baseline \, fitting over a width of
\begin{landscape}  
\begin{table}[t]
 \normalsize
 \renewcommand{\baselinestretch}{0.85}
\caption{\HI\ non-detections in the Hydra/Antlia region}
\label{ndet}
\scriptsize  
\begin{tabular*}{22cm}{
  l  @{\extracolsep{4mm}} p{1.4cm} @{\extracolsep{1mm}}
  c @{\extracolsep{2mm}}                                                                                                                                                   
  l@{\extracolsep{3mm}} l @{\extracolsep{4mm}}
  r @{\extracolsep{2mm}}r @{\extracolsep{4mm}} 
  p{2.7mm} @{\extracolsep{-1mm}}
  p{2.7mm} @{\extracolsep{-1mm}}
  p{2.7mm} @{\extracolsep{-1mm}} 
  p{2.7mm} @{\extracolsep{0mm}}
  p{6mm} @{\extracolsep{0mm}} p{4.5mm} @{\extracolsep{3mm}}
  r @{\extracolsep{2mm}} c @{\extracolsep{2mm}} 
  r @{\extracolsep{4mm}} r @{\extracolsep{4mm}} 
%
 r @{\extracolsep{2mm}} c @{\extracolsep{2mm}} r @{\extracolsep{2mm}} 
 r @{\extracolsep{2mm}}                        
 r @{\extracolsep{2mm}} c @{\extracolsep{2mm}} r @{\extracolsep{4mm}} 
 r @{\extracolsep{0mm}}                        
}
\hline
\vspace{-1mm} \\
 Ident. & Other & IR & \ \ \ \ R.A. & \ \ \ Dec.& gal $\ell$ \ & gal $b$ \ &
 \multicolumn{4}{l}{Type} & 
 \multicolumn{2}{c}{$D$ x $d$} & 
 $B_{J}$ & 
 $E_{(B-V)}$ & 
 $D_{J}^0$ &
 $B_{J}^0$ &
 \multicolumn{3}{c}{$V_{range}^{obs}$} & 
 {$rms$} &
 \multicolumn{3}{c}{$V_{range}^{pert.}$} & 
 $V_{other}$ \\ 
& &  &
(h\,\, m\,\, s) & ($\deg$\,\, $\arcmin$\,\,$\arcsec$) & ($\deg$) \ &($\deg$) \ &
& & & &
\multicolumn{2}{c}{($\arcsec$)} & ($^{\rm m}$) & ($^{\rm m}$) &
($\arcsec$) & ($^{\rm m}$) & 
 \multicolumn{3}{c}{km/s} & m\,Jy & \multicolumn{3}{c}{km/s} &  km/s \\
\vspace{-1mm} \\
\ \ \ (1) & \ \ \ (2) & (3) & \ \ \ \ (4) & \ \ \ \ (5) & (6) \ & (7) \
 & \multicolumn{4}{c}{(8)} \ & \multicolumn{2}{c}{(9)} \ & \ \ (10) &
 {(11)} & (12) & (13) &  \multicolumn{3}{c}{(14)} &  (15) &  \multicolumn{3}{c}{(16)}  & (17) \ \ \\
\hline
\vspace{-1mm} \\
RKK0142 &          &    & 08 36 48.2 & -56 28 55 &  272.83 &  -9.33 & S&B& &5& \hfill  40x&\hfill  27 & 16.4 & 0.19 &  47 & 15.6 &  500& -- & 13000 & 4.0 &  8100& -- & 8300 &             \\ 
RKK0230 &          &    & 08 38 35.5 & -58 03 52 &  274.27 & -10.07 & S& & &M& \hfill  51x&\hfill  27 & 16.2 & 0.17 &  58 & 15.5 &  400& -- & 12800 & 3.5 &      &    &      &             \\ 
RKK0349 &          &    & 08 42 42.7 & -53 55 26 &  271.27 &  -7.12 & S& & &M& \hfill  67x&\hfill   7 & 17.3 & 0.38 & 112 & 15.5 &  600& -- & 12800 & 3.3 &  1100& -- & 1300 &             \\ 
RKK0385 &          &    & 08 43 53.4 & -56 34 31 &  273.50 &  -8.60 & S&B& &5& \hfill  40x&\hfill  40 & 16.0 & 0.20 &  48 & 15.1 &  600& -- & 10000 & 3.7 &  7900& -- & 8400 &  13786 (M)  \\ 
RKK0476 &          &    & 08 47 17.1 & -54 26 25 &  272.10 &  -6.91 & S& & & & \hfill  34x&\hfill  34 & 16.4 & 0.37 &  54 & 14.7 &  600& -- &  5600 & 3.2 &      &    &      &             \\ 
RKK0566 &          &    & 08 50 33.9 & -59 58 54 &  276.77 & -10.00 & S& & &M& \hfill  60x&\hfill  13 & 16.9 & 0.18 &  69 & 16.1 &  600& -- & 10000 & 3.6 &  7900& -- & 8400 &  23980 (M)  \\ 
RKK0654 & L165-006 &    & 08 53 36.3 & -54 05 07 &  272.40 &  -5.97 & D& & & & \hfill 134x&\hfill 121 & 13.7 & 0.34 & 216 & 12.1 &  600& -- & 10000 & 3.4 &      &    &      &             \\ 
RKK0677 &          &    & 08 54 06.4 & -58 20 55 &  275.77 &  -8.62 & S& & &5& \hfill  34x&\hfill  28 & 16.4 & 0.25 &  44 & 15.3 &  500& -- &  9500 & 3.0 &      &    &      &             \\ 
RKK0775 &          &    & 08 57 56.8 & -59 17 15 &  276.82 &  -8.84 & S& & &M& \hfill  30x&\hfill  27 & 16.5 & 0.23 &  37 & 15.5 &  600& -- & 10000 & 3.9 &      &    &      &             \\ 
RKK0863 &          &    & 09 01 12.2 & -60 00 21 &  277.65 &  -8.99 & S& & &5& \hfill  32x&\hfill  30 & 16.4 & 0.19 &  37 & 15.6 &  600& -- &  5600 & 3.3 &      &    &      &             \\ 
\vspace{-1.00mm} \\ 
RKK0889 &          &    & 09 02 12.4 & -58 54 48 &  276.89 &  -8.18 & S& & &1& \hfill  62x&\hfill  28 & 15.9 & 0.30 &  87 & 14.5 &  600& -- & 13000 & 2.9 &      &    &      &  10361 (1)  \\ 
RKK0902 &          &    & 09 02 57.5 & -53 29 10 &  272.85 &  -4.53 & S& & & & \hfill  40x&\hfill  11 & 17.3 & 0.83 & 254 & 12.4 &  700& -- & 10000 & 3.9 &      &    &      &             \\ 
RKK0906 &          &    & 09 02 58.2 & -54 08 44 &  273.35 &  -4.97 & S& & & & \hfill  51x&\hfill   8 & 17.1 & 0.50 & 103 & 14.8 &  600& -- &  5600 & 3.1 &      &    &      &             \\ 
RKK0966 &          & I  & 09 05 13.0 & -54 26 30 &  273.79 &  -4.92 & S& & &L& \hfill  40x&\hfill   7 & 17.5 & 0.48 &  82 & 15.2 &  600& -- & 10000 & 3.2 &      &    &      &             \\ 
RKK0972 &          &    & 09 05 29.8 & -55 00 48 &  274.49 &  -5.49 & S& & & & \hfill  54x&\hfill  12 & 16.6 & 0.37 &  85 & 14.9 &  500& -- & 10000 & 4.6 &      &    &      &             \\ 
RKK0980 &          &    & 09 05 41.8 & -55 48 50 &  274.86 &  -5.78 & D& & & & \hfill  54x&\hfill  40 & 15.8 & 0.35 &  88 & 14.2 &  400& -- & 10000 & 3.6 &      &    &      &             \\ 
RKK0986 &          &    & 09 05 52.3 & -55 22 55 &  274.55 &  -5.47 & S& & &1& \hfill  40x&\hfill   9 & 17.1 & 0.37 &  64 & 15.4 & 6000& -- & 11100 & 3.0 &      &    &      &   9832 (1)  \\ 
RKK1041 &          &    & 09 08 55.1 & -54 59 18 &  274.55 &  -4.89 & S& & &5& \hfill  27x&\hfill  20 & 17.3 & 0.50 &  61 & 14.7 &  500& -- &  5600 & 2.8 &      &    &      &             \\ 
RKK1180 &          &    & 09 16 20.5 & -63 12 52 &  281.28 &  -9.79 &  & & & & \hfill  40x&\hfill  27 & 15.9 & 0.20 &  47 & 15.0 &  600& -- & 12800 & 3.6 &      &    &      &             \\ 
RKK1194 & L126-005 &    & 09 17 41.0 & -59 36 58 &  278.76 &  -7.19 & S& & &4& \hfill  60x&\hfill  27 & 15.5 & 0.28 &  81 & 14.2 &  500& -- & 12000 & 3.4 &      &    &      &  10240 (2)  \\ 
\vspace{-1.00mm} \\ 
RKK1204 &          &    & 09 18 12.1 & -56 40 02 &  276.68 &  -5.09 & S& & &5& \hfill  27x&\hfill  12 & 17.8 & 0.61 &  75 & 14.8 &  700& -- &  5600 & 2.8 &      &    &      &             \\ 
RKK1235 &          &    & 09 19 39.0 & -63 22 10 &  281.66 &  -9.63 & S& & & & \hfill  47x&\hfill  27 & 16.7 & 0.22 &  57 & 15.7 &  500& -- & 10000 & 4.0 &      &    &      &             \\ 
RKK1274 &          &    & 09 22 43.3 & -63 21 52 &  281.90 &  -9.38 & S& & &4& \hfill  34x&\hfill  13 & 17.2 & 0.24 &  43 & 16.1 &  600& -- & 10000 & 3.7 &  7900& -- & 8400 &             \\ 
RKK1284$^a$&          &    & 09 23 18.7 & -63 49 51 &  282.28 &  -9.66 & S& & & & \hfill  40x&\hfill   7 & 17.5 & 0.23 &  49 & 16.5 &  600& -- & 10000 & 2.4 &  1100& -- & 1500 &             \\ 
        &          &    &            &           &         &        &  & & & & \hfill     &\hfill     &      &      &     &      &     &    &       &     &  7900& -- & 8200 &             \\ 
RKK1369 &          &    & 09 28 17.1 & -61 53 31 &  281.31 &  -7.88 & S& & &5& \hfill  54x&\hfill  34 & 16.2 & 0.31 &  79 & 14.7 &  500& -- & 10000 & 3.6 &      &    &      &             \\ 
RKK1409 &          & I  & 09 30 49.4 & -58 03 30 &  278.87 &  -4.89 & S& & &M& \hfill  34x&\hfill   5 & 18.0 & 0.67 & 121 & 14.4 &  800& -- &  9700 & 3.9 &      &    &      &             \\ 
RKK1411 &          & I  & 09 30 55.3 & -57 36 52 &  278.58 &  -4.56 & S& & & & \hfill  34x&\hfill  27 & 16.4 & 0.91 & 283 & 10.9 &  300& -- & 10000 & 4.1 &      &    &      &             \\ 
RKK1457 &          &    & 09 33 43.1 & -62 40 56 &  282.32 &  -8.02 & S& & & & \hfill  47x&\hfill  40 & 15.9 & 0.30 &  66 & 14.5 &  600& -- &  5600 & 2.5 &      &    &      &             \\ 
RKK1470 &          &    & 09 35 06.4 & -62 20 18 &  282.20 &  -7.66 & S& & &E& \hfill  30x&\hfill  11 & 17.4 & 0.34 &  44 & 15.9 &  600& -- & 10000 & 3.2 &  8000& -- & 8300 &             \\ 
\vspace{-1.00mm} \\ 
RKK1482 &          &    & 09 36 10.7 & -62 23 60 &  282.34 &  -7.62 & S& & &3& \hfill  60x&\hfill  54 & 15.2 & 0.31 &  85 & 13.8 &  600& -- & 10000 & 2.6 &      &    &      &             \\ 
RKK1543 &          &    & 09 40 44.8 & -49 31 37 &  274.23 &   2.41 & S& & & & \hfill  31x&\hfill   7 & 17.8 & 0.67 & 115 & 14.1 &  600& -- & 10000 & 3.6 &  4250& -- & 4600 &             \\ 
RKK1554 &          &    & 09 41 41.2 & -49 33 07 &  274.36 &   2.50 & S&B& &5& \hfill  34x&\hfill  27 & 16.5 & 0.75 & 153 & 12.4 &  600& -- &  5600 & 2.4 &      &    &      &             \\ 
RKK1621 &          &    & 09 46 05.5 & -49 21 17 &  274.77 &   3.11 & S&B& &5& \hfill  24x&\hfill  20 & 17.2 & 0.53 &  59 & 14.4 &  600& -- &  5600 & 2.4 &      &    &      &             \\ 
RKK1669 &          &    & 09 49 27.4 & -49 08 26 &  275.06 &   3.63 & S& & &4& \hfill  34x&\hfill  34 & 16.0 & 0.54 &  87 & 13.2 &  600& -- & 10200 & 3.1 &      &    &      &             \\ 
RKK1707 &          &    & 09 51 43.0 & -47 21 59 &  274.22 &   5.24 & I& & & & \hfill  81x&\hfill  13 & 16.1 & 0.41 & 143 & 14.1 &  600& -- & 10000 & 3.6 &  1000& -- & 1300 &             \\ 
RKK1749 &          &    & 09 54 46.9 & -49 45 53 &  276.13 &   3.69 & S& & &L& \hfill  47x&\hfill   5 & 17.6 & 0.38 &  76 & 15.9 &  800& -- &  5600 & 2.7 &      &    &      &             \\ 
RKK1767 &          &    & 09 55 16.  & -61 57.2  &  283.75 &  -5.83 & S& & &4& \hfill  27x&\hfill  20 & 16.4 & 0.35 &  42 & 14.8 &  600& -- &  5400 & 2.2 &      &    &      &             \\ 
RKK1780 & FGCE0780 &    & 09 55 31.0 & -67 07 54 &  287.04 &  -9.86 & S& & & & \hfill  71x&\hfill   8 & 16.8 & 0.23 &  87 & 15.8 &  600& -- & 10000 & 4.2 &      &    &      &             \\ 
RKK1894 &          &    & 10 03  6.5 & -53 15 34 &  279.29 &   1.70 & I& & & & \hfill  24x&\hfill  20 & 17.0 & 0.90 & 207 & 11.4 &  400& -- &  5300 & 2.1 &  1100& -- & 1400 &  17688 (M)  \\ 
\vspace{-1.00mm} \\ 
RKK1900 &          &    & 10 03 26.4 & -50 05 16 &  277.43 &   4.28 & S& & &9& \hfill  38x&\hfill   4 & 17.8 & 0.47 &  79 & 15.4 &  600& -- &  5500 & 2.2 &      &    &      &             \\ 
RKK1910 &          &    & 10 03 43.5 & -66 12 04 &  287.11 &  -8.63 & S& & &M& \hfill  40x&\hfill  12 & 16.8 & 0.22 &  49 & 15.8 &  600& -- & 10000 & 3.3 &      &    &      &             \\ 
RKK1963 & L092-010 &  I & 10 07 00.2 & -64 21 47 &  286.28 &  -6.95 & S& & &3& \hfill  74x&\hfill  74 & 14.3 & 0.22 &  90 & 13.3 &  300& -- &  5000 & 3.3 &      &    &      &     57 (1)  \\ 
RKK1995 &          &    & 10 08 42.8 & -61 26 19 &  284.72 &  -4.46 & S& & & & \hfill  34x&\hfill  20 & 16.3 & 0.37 &  55 & 14.5 &  600& -- & 10000 & 3.4 &  7000& -- & 7200 &             \\ 
RKK2017 & HEN373   & I  & 10 10 02.5 & -57 01 54 &  282.30 &   -.77 & S& & & & \hfill  67x&\hfill  27 & 15.8 & 6.30 & 999 &  3.6 &  600& -- & 10000 & 4.5 &  7500& -- & 8800 &             \\ 
RKK2095 &          & I  & 10 13 25.5 & -52 31 21 &  280.12 &   3.21 & S& & &3& \hfill  40x&\hfill  20 & 16.0 & 0.54 &  99 & 13.2 &  400& -- &  5400 & 2.8 &      &    &      &   8576 (2)  \\ 
RKK2118 &          &    & 10 13 43.3 & -62 21 05 &  285.73 &  -4.87 & S& & &5& \hfill  54x&\hfill   5 & 17.6 & 0.35 &  82 & 16.0 &  500& -- & 10000 & 3.7 &      &    &      &             \\ 
RKK2142 &          &    & 10 14 53.4 & -47 43 52 &  277.59 &   7.29 & S&B& &4& \hfill  40x&\hfill  20 & 16.2 & 0.19 &  47 & 15.4 &  600& -- & 10000 & 3.3 &      &    &      &             \\ 
RKK2154 &          &    & 10 14 58.3 & -62 22 05 &  285.86 &  -4.80 & S& & & & \hfill  40x&\hfill  34 & 16.2 & 0.37 &  63 & 14.5 &  600& -- & 10000 & 2.6 &      &    &      &             \\ 
RKK2157 &          &    & 10 15 28.3 & -51 19 23 &  279.71 &   4.38 & S& & &M& \hfill  34x&\hfill  23 & 16.1 & 0.47 &  74 & 13.6 &  600& -- &  5600 & 2.8 &      &    &      &  14078 (M)  \\ 
\vspace{-1.00mm} \\ 
RKK2168 &          &    & 10 15 31.4 & -61 47 32 &  285.58 &  -4.29 & S& & &L& \hfill  54x&\hfill  11 & 16.5 & 0.43 & 102 & 14.4 &  400& -- &  5600 & 2.3 &      &    &      &             \\ 
RKK2173 &          &    & 10 16 03.5 & -52 10 11 &  280.26 &   3.73 & S& & & & \hfill  40x&\hfill   5 & 18.2 & 0.46 &  81 & 15.9 &  500& -- &  9500 & 2.9 &  9500& -- &10200 &             \\ 
RKK2198 &          &    & 10 16 45.8 & -53 31 02 &  281.10 &   2.67 & S& & & & \hfill  27x&\hfill   9 & 17.9 & 0.64 &  90 & 14.5 &  600& -- & 10200 & 3.2 &  1100& -- & 1600 &             \\ 
        &          &    &            &           &         &        &  & & & & \hfill     &\hfill     &      &      &     &      &     &    &       &     &  7400& -- & 7600 &             \\ 
RKK2310 &          & I  & 10 19 25.8 & -60 59 18 &  285.53 &  -3.36 & E& & & & \hfill  54x&\hfill  24 & 15.2 & 0.79 & 216 & 11.3 &  600& -- & 10000 & 3.6 &      &    &      &             \\ 
RKK2317 &          & I  & 10 19 53.3 & -47 47 28 &  278.33 &   7.71 & S& & &M& \hfill  40x&\hfill  22 & 16.4 & 0.15 &  44 & 15.8 &  600& -- & 10000 & 2.3 &      &    &      &  17819 (3)  \\ 
RKK2337 &          &    & 10 20 27.7 & -49 22 54 &  279.29 &   6.43 & I& & &M& \hfill  59x&\hfill  40 & 15.6 & 0.26 &  77 & 14.4 &  600& -- & 13000 & 3.6 &      &    &      &             \\ 
RKK2361 &          &    & 10 21 11.5 & -50 45 30 &  280.14 &   5.34 & S& & &M& \hfill  34x&\hfill  24 & 16.4 & 0.34 &  52 & 14.8 &  500& -- &  5500 & 2.2 &      &    &      &             \\ 
RKK2380 &          & I  & 10 21 40.8 & -49 36 53 &  279.58 &   6.35 &  & & & & \hfill  40x&\hfill  30 & 16.3 & 0.23 &  50 & 15.3 &  600& -- & 10000 & 3.4 &  7800& -- & 8400 &  14741 (3)  \\ 
RKK2391 &          &    & 10 22 01.0 & -49 35 53 &  279.62 &   6.39 &  & & & & \hfill  44x&\hfill  16 & 17.4 & 0.24 &  55 & 16.4 &  800& -- & 10000 & 3.6 &  1200& -- & 1300 &             \\ 
\vspace{-1.00mm} \\ 
\hline
 \end{tabular*}
 \normalsize
 \end{table}
\addtocounter{table}{-1}
\clearpage
\begin{table}[t]
 \normalsize
 \renewcommand{\baselinestretch}{0.85}
\caption{\HI\ non-detections in the Hydra/Antlia region -- (continued)}
\scriptsize  
\begin{tabular}{
  l  @{\extracolsep{4mm}} p{1.4cm} @{\extracolsep{1mm}}
  c @{\extracolsep{2mm}}                                                                                                                                                   
  l@{\extracolsep{3mm}} l @{\extracolsep{4mm}}
  r @{\extracolsep{2mm}}r @{\extracolsep{4mm}} 
  p{2.7mm} @{\extracolsep{-1mm}}
  p{2.7mm} @{\extracolsep{-1mm}}
  p{2.7mm} @{\extracolsep{-1mm}} 
  p{2.7mm} @{\extracolsep{0mm}}
  p{6mm} @{\extracolsep{0mm}} p{4.5mm} @{\extracolsep{3mm}}
  r @{\extracolsep{2mm}} c @{\extracolsep{2mm}} 
  r @{\extracolsep{4mm}} r @{\extracolsep{4mm}} 
%
 r @{\extracolsep{2mm}} c @{\extracolsep{2mm}} r @{\extracolsep{2mm}} 
 r @{\extracolsep{2mm}}                        
 r @{\extracolsep{2mm}} c @{\extracolsep{2mm}} r @{\extracolsep{4mm}} 
 r @{\extracolsep{0mm}}                        
}
\hline
\vspace{-1mm} \\
 Ident. & Other & IR & \ \ \ \ R.A. & \ \ \ Dec.& gal $\ell$ \ & gal $b$ \ &
 \multicolumn{4}{l}{Type} & 
 \multicolumn{2}{c}{$D$ x $d$} & 
 $B_{J}$ & 
 $E_{(B-V)}$ & 
 $D_{J}^0$ &
 $B_{J}^0$ &
 \multicolumn{3}{c}{$V_{range}^{obs}$} & 
 {$rms$} &
 \multicolumn{3}{c}{$V_{range}^{pert.}$} & 
 $V_{other}$ \\ 
& &  &
(h\,\, m\,\, s) & ($\deg$\,\, $\arcmin$\,\,$\arcsec$) & ($\deg$) \ &($\deg$) \ &
& & & &
\multicolumn{2}{c}{($\arcsec$)} & ($^{\rm m}$) & ($^{\rm m}$) &
($\arcsec$) & ($^{\rm m}$) & 
 \multicolumn{3}{c}{km/s} & m\,Jy & \multicolumn{3}{c}{km/s} &  km/s \\
\vspace{-1mm} \\
\ \ \ (1) & \ \ \ (2) & (3) & \ \ \ \ (4) & \ \ \ \ (5) & (6) \ & (7) \
 & \multicolumn{4}{c}{(8)} \ & \multicolumn{2}{c}{(9)} \ & \ \ (10) &
 {(11)} & (12) & (13) &  \multicolumn{3}{c}{(14)} &  (15) &  \multicolumn{3}{c}{(16)}  & (17) \ \ \\
\hline
\vspace{-1mm} \\
RKK2445 &          &    & 10 23 50.5 & -48 12 25 &  279.12 &   7.72 & S& &R&2& \hfill  34x&\hfill  34 & 16.2 & 0.16 &  38 & 15.5 &  600& -- &  5500 & 2.6 &      &    &      &             \\ 
RKK2459 &          &    & 10 24 16.6 & -49 54 27 &  280.10 &   6.33 & S& & & & \hfill  42x&\hfill   7 & 17.8 & 0.32 &  61 & 16.3 &  700& -- & 10000 & 3.3 &  7900& -- & 8300 &             \\ 
RKK2525$^a$& L214-005 &  I & 10 26 54.7 & -49 08 53 &  280.06 &   7.20 & S& & &3& \hfill  47x&\hfill  20 & 16.2 & 0.29 &  64 & 14.9 &  600& -- & 12800 & 6.2 &  3600& -- & 4600&   14450 \ (6)\\
RKK2512 &          & Q  & 10 26 10.3 & -53 46 55 &  282.42 &   3.20 & F& & & & \hfill  27x&\hfill  27 & 16.1 & 0.50 &  60 & 13.6 &  600& -- &  5600 & 3.7 &      &    &      &             \\ 
RKK2549 &          &    & 10 27 24.2 & -50 28 34 &  280.83 &   6.11 & S& & &5& \hfill  38x&\hfill  22 & 16.7 & 0.26 &  49 & 15.6 &  400& -- & 10200 & 3.6 &  1200& -- & 1600 &  15934 (M)  \\ 
RKK2588 &          & I  & 10 28 05.3 & -64 29 15 &  288.25 &  -5.78 & S& & & & \hfill  39x&\hfill  20 & 16.9 & 0.66 & 151 & 13.1 &  400& -- & 10000 & 3.2 &  1900& -- & 2250 &             \\ 
RKK2592 &          &    & 10 28 40.8 & -49 28 48 &  280.48 &   7.07 &  & & & & \hfill  47x&\hfill  34 & 15.6 & 0.33 &  69 & 14.1 &  500& -- & 12800 & 2.9 &      &    &      &  17975 (M)  \\ 
RKK2598$^a$& L214-011 &    & 10 28 48.1 & -50 41 57 &  281.14 &   6.04 &
S& & &1& \hfill  67x&\hfill  32 & 15.4 & 0.29 &  91 & 14.1 & 4000& --
& 12800 & 2.7 &      &    &      &   6305 \ (1)  \\ 
RKK2603 &          &    & 10 29  3.3 & -51 41 21 &  281.69 &   5.21 & S& & &M& \hfill  38x&\hfill  27 & 16.4 & 0.25 &  49 & 15.3 &  600& -- & 10000 & 3.9 &      &    &      &  19557 (M)  \\ 
RKK2649 &          & P  & 10 30 54.4 & -51 22 20 &  281.77 &   5.63 & S& & &M& \hfill  36x&\hfill  27 & 16.4 & 0.40 &  63 & 14.5 &  800& -- &  5500 & 4.3 &      &    &      &             \\ 
\vspace{-1.00mm} \\ 
RKK2663$^a$&          &    & 10 31 28.9 & -63 44 07 &  288.18 &  -4.95 & S& & &L& \hfill  50x&\hfill   8 & 17.4 & 0.47 & 104 & 15.0 & 1000& -- & 10000 & 8.5 & (7300& -- & 8500)&             \\ 
RKK2743 &          &    & 10 36 05.2 & -50 14 16 &  281.90 &   7.02 & S& & &M& \hfill  31x&\hfill  28 & 16.4 & 0.41 &  52 & 14.6 &  600& -- & 10200 & 2.8 &  1200& -- & 1600 &             \\ 
        &          &    &            &           &         &        &  & & & & \hfill     &\hfill     &      &      &     &      &     &    &       &     &  7400& -- & 7700 &             \\ 
        &          &    &            &           &         &        &  & & & & \hfill     &\hfill     &      &      &     &      &     &    &       &     &  9800& -- &10000 &             \\ 
RKK2774 &          & I  & 10 39 00.8 & -50 25 18 &  282.40 &   7.10 & S&B& &3& \hfill  43x&\hfill  31 & 15.7 & 0.55 & 107 & 12.9 &  600& -- &  5600 & 3.2 &  1200& -- & 1400 &   7181 (4)  \\ 
RKK2778 &          &    & 10 39 15.6 & -52 26 16 &  283.43 &   5.36 & S& & & & \hfill  40x&\hfill  20 & 16.6 & 0.44 &  75 & 14.5 &  500& -- & 10000 & 4.3 &  7000& -- & 7200 &             \\ 
RKK2784 &          &    & 10 39 44.2 & -50 32 06 &  282.56 &   7.05 & S& &R&M& \hfill  43x&\hfill  40 & 16.2 & 0.53 & 103 & 13.5 &  500& -- &  5600 & 2.6 &      &    &      &             \\ 
RKK2797 &          &    & 10 40 34.4 & -49 25 01 &  282.13 &   8.10 & I& & &9& \hfill  42x&\hfill  31 & 16.1 & 0.38 &  69 & 14.3 &  400& -- &  5600 & 2.8 &      &    &      &             \\ 
RKK2807 &          &    & 10 41 31.6 & -51 32 32 &  283.30 &   6.31 & S& & & & \hfill  43x&\hfill   9 & 17.5 & 0.44 &  82 & 15.4 &  400& -- & 10000 & 3.7 &      &    &      &             \\ 
RKK2841 &          &    & 10 47 19.5 & -50 28 56 &  283.60 &   7.67 & S& & &2& \hfill  47x&\hfill  15 & 16.6 & 0.42 &  86 & 14.6 &  400& -- &  9900 & 2.9 &  1100& -- & 1300 &             \\ 
        &          &    &            &           &         &        &  & & & & \hfill     &\hfill     &      &      &     &      &     &    &       &     &  1900& -- & 2100 &             \\ 
\vspace{-1.00mm} \\ 
RKK2845 &          &    & 10 48 07.7 & -50 29 47 &  283.73 &   7.72 & S& & &L& \hfill  47x&\hfill  17 & 16.5 & 0.43 &  89 & 14.4 &  600& -- &  5600 & 2.4 &      &    &      &             \\ 
RKK2895 &          & I  & 11 10 05.9 & -67 23 40 &  293.40 &  -6.42 & S&B& &3& \hfill  74x&\hfill  40 & 15.6 & 0.35 & 114 & 14.0 &  600& -- & 12800 & 4.5 &      &    &      &             \\ 
RKK2896 &          & Q  & 11 12 10.1 & -57 33 11 &  289.90 &   2.79 & S& & & & \hfill  30x&\hfill   3 & 18.8 & 0.86 & 229 & 13.5 &  800& -- &  5600 & 2.0 &      &    &      &             \\ 
RKK2931 &          &    & 11 20 21.9 & -59 10 26 &  291.49 &   1.66 & S& & & & \hfill  40x&\hfill  11 & 17.0 & 0.79 & 213 & 12.5 &  400& -- & 10000 & 4.2 &      &    &      &             \\ 
RKK2967 &          &    & 11 24 17.4 & -56 17 06 &  291.00 &   4.56 & S& & &M& \hfill  74x&\hfill  40 & 15.3 & 0.41 & 126 & 13.4 &  400& -- & 10300 & 2.9 &  1000& -- & 1200 &             \\ 
RKK2976 &          &    & 11 24 51.5 & -57 11 15 &  291.38 &   3.73 & S& & & & \hfill  74x&\hfill  60 & 15.1 & 0.38 & 120 & 13.3 &  600& -- & 12800 & 3.7 &      &    &      &             \\ 
\vspace{-1.00mm} \\ 
\hline
\end{tabular}
 \normalsize
\label{nondet}
\end{table}
\end{landscape}

\noindent 
1600\kms\ -- hence similar to the
determination for detections -- centered at increasingly higher
redshifts in order to obtain values for the whole velocity range. The
quoted values represent the highest rms for the velocity
intervals, the rms for the nearer velocities are on average a bit
lower.

{\bf Col. 16:} Perturbed velocity intervals, mainly due to recurring
rfi around 1250 and 4450 \kms\ and very strong GPS signals around
8300\kms. In these intervals a signal would have gone unnoticed.
Narrow interference patterns that are not likely to hide the \HI\
emission of a galaxy are not marked in this column.

{\bf Col. 17:} Independent velocity determinations for the
non-detected galaxies. The reference coding is the same as in
Table~\ref{det} with the addition of M, which stands for observations
made with the multifiber spectrograph MEFOS at the 3.6\,m telescope of
ESO which will be published shortly (Woudt \etal 2002).  These
velocities reveal that the non-detections are either beyond the
covered velocity range or so distant that their signals would be too weak
for detection with the achieved noise level.

\subsection{Signals in the spectra of non-detectons} \label{ndetsig}
Some of the spectra of the non-detections listed in Table~{\ref{ndet}
do reveal a positive signal. Comparing the recession velocities with
independent optical data proves, however, that the observed \HI\ emission
in four of the pointings does not originate from the targeted object but
from a neighbour at close angular distance. This concerns the
following spectra:

{\bf RKK\,1284:} At this pointing the \HI\ emission dominating the
spectrum stems from the nearby ($\Delta r = 9\farcm4$) large spiral
galaxy RKK\,1288 = ESO\,091-011 at 3197\kms, as confirmed by its
optically determined redshift of 3192\kms\ (see column 19 in
Table~\ref{det}). However, underneath the broad signal ($\Delta
V_{\rm 20} = 306$\kms) there is an indication of a narrower signal
($\Delta V_{\rm 20} = 117$\kms) at a slightly lower velocity (3117\kms).
Synthesis observations are required to confirm whether this signal is
real and due to RKK\,1284.

{\bf RKK\,2525:} Our observations find a weak, although clear
\HI\ detection at 5482\kms\ which was confirmed in a repeat
observation. However, RKK\,2525 has a velocity listed in the 1992 NED
compilation (without a source) of 14\,450\kms\ which is independently
confirmed by our multifiber spectroscopy observations obtained with
MEFOS at the 3.6~m telescope of ESO (Woudt \etal 2002).  The most
likely object for the observed \HI\ emission is the fairly edge-on
galaxy RKK\,2546 at $\Delta r = 4\farcm3$ which has an optical
velocity of $V_{\rm hel} = 5492$\kms\ (Kraan-Korteweg \etal 1995).

{\bf RKK\,2598} (= ESO\,214-11): The signal identified at this
pointing is a weaker replica of the profile found for the galaxy
RKK\,2585 at a separation of $\Delta r = 4\farcm0$ (see Table~1 and
Fig.~\ref{profile}.  The velocity of RKK\,2585 has been confirmed in
the optical ($V_{\rm hel} = 6868$\kms) whereas the optical velocity for
RKK\,2598 is 6305\kms, considerably lower than the \HI\ measurement
(7027\kms).  There is no signal in the spectrum near the optical
redshift even though the galaxy is quite large, but it is classified
as an Sa.

{\bf RKK\,2663} is a similar case. The detection at $V$ = 3759\kms\ of
$I$ = 6.38\,Jy\kms\ is due to the galaxy RKK\,2684 = ESO\,092-019 also
detected here (see Table~1 and Fig.~\ref{profile}) and its origin is
confirmed by the optical velocity of the latter.
 
\section{Description of the observed sample}

We observed mainly extended, LSB
galaxies -- either galaxies that are intrinsically LSB or galaxies
reduced in size and surface brightness due to the heavy foreground
obscuration, for which optical spectroscopy will not yield good S/N
spectra. Furthermore, observations were made of nearby
spiral galaxies with known optical redshifts but for which no prior
\HI\ observations existed. These were selected such that all spiral
galaxies with an extinction-corrected diameter $D^{\rm o} \ga 60\arcsec$
will have been observed in the 21\,cm line. The resulting \HI\ sample
forms part of our project to map the peculiar velocity field in the
ZOA via the near infrared Tully--Fisher relation.

\begin{figure}[t]
\epsfig{file=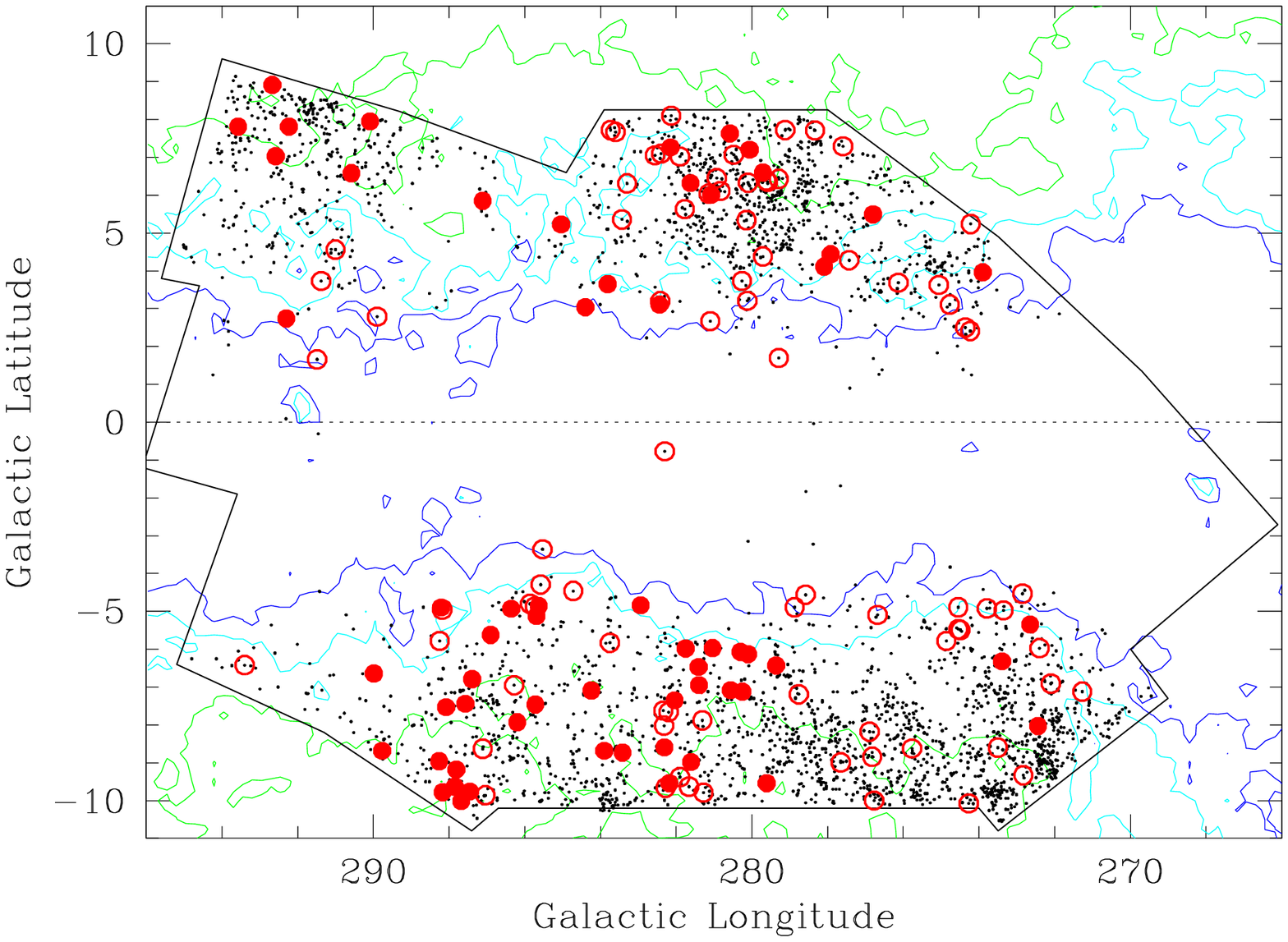,width=8.8cm} 
\caption{Distribution of the galaxies observed at 21-cm in the
Hydra/Antlia ZOA region. Filled circles indicate the 66 detections,
open circles the 82 non-detections. The small dots represent the
galaxies uncovered within the optical search region (outlined area)
and the contours show the dust extinction as determined from the 100$\mu$m
DIRBE maps (Schlegel \etal 1998) at the levels $A_{\rm B} = 1\fm0$,
2$\fm$0 and 3$\fm$0 (thick line).}
\label{dist}
\end{figure}

Fig.~\ref{dist} shows the Hydra/Antlia search area (outlined region)
with the 3279 optically discovered galaxies with $D \la 0\fm2$ plotted as small
dots. The displayed extinction contours are equivalent to absorption
levels in the blue of $A_{\rm B} = 1\fm$0, 2$\fm$0 and 3$\fm$0. The
figure demonstrates the effectiveness of optical galaxy searches in
reducing the ZOA and revealing the underlying large-scale distribution
of galaxies at least to extinction levels of $A_{\rm B} = 3\fm0$ (at the highest
extinction levels, only blind \HI\ surveys are successful in tracing
large-scale structure in the nearby Universe).  
The 139 galaxies observed with the
Parkes radio telescope are indicated with big symbols: filled circles
illustrate detections and open circles the non-detections. An
inspection of the pointings within the survey region show them to be
distributed quite homogeneously over the galaxy distribution.

There are clear discrepancies, however, between the distribution of
the detections compared to the non-detections. These can be completely
attributed to different large-scale structures (LSS) crossing the
Galactic Plane (GP) in this region. As found from the individual
(Fairall \etal 1998) and multifibre spectroscopy (Kraan-Korteweg \etal
1994, Woudt \etal 2002), the concentration of galaxies below the GP in
the longitude range $270\deg \la \ell \la 280\deg$ is primarily due
to clustering at high redshifts, i.e. at $9 - 11\,000$ and $16 -
18\,000$\,\kms\ (see also column 17, Table~\ref{ndet}), hence beyond
our \HI\ velocity search range and sensitivity limit. Similarly, the
scarcity of detections in the dense region above the Galactic equator
around $\ell \sim 280\deg$ can be explained by the fact the we 
expect to detect only spiral galaxies related to the Hydra/Antlia
filament ($V\sim 3000$\,\kms) and the \HI-richest galaxies belonging
to the Vela supercluster ($V\sim 6000$\kms), but none of the
galaxies at about 16\,000\kms\ which also form part of that
condensation.

Besides the effects of large-scale structure in the distribution of
detections versus non-detections, Fig.~\ref{dist} seems to support the
interpretation that the \HI\ detection rate is independent of latitude,
and thus, of extinction. An analysis of the detection versus
non-detection rate as a function of extinction (not shown here)
confirms this.

\begin{figure}[t]
\epsfig{file=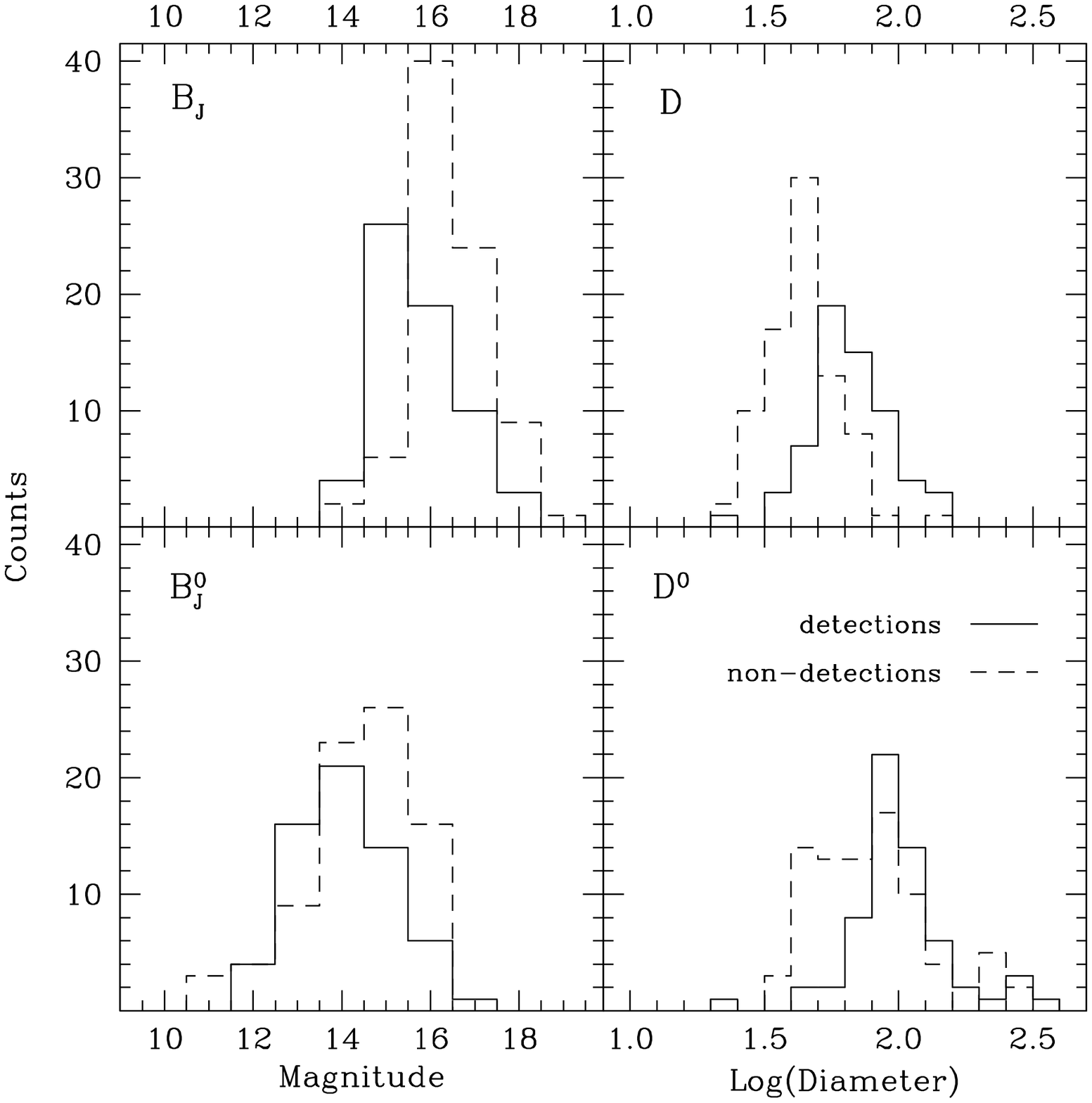,width=8.cm}
    \caption{Histograms of the detections (bold line) and
    non-detections (dashed line) for observed (upper panels) and
    extinction-corrected (lower panels) magnitudes (left) and
    diameters (right).  }
\label{allhist}
\end{figure}

Figure~\ref{allhist} illustrates the typical
properties of the observed galaxy sample. The top panels show the
distribution of the observed magnitudes ($B_{\rm J}$) and diameters ($\log
D$) divided into detections (bold line) and non-detections (dashed
line). We can see a clear offset in the distribution of the detections
($<B_{\rm J}> = 15\fm74$ and $<D> = 68\farcs6$) compared to the
non-detections ($<B_{\rm J}> = 16\fm62$ and $<D> = 44\farcs2$) with the
detections being nearly 1 magnitude brighter and typically 1 1/2 times
larger. This is an expected trend for observations unaffected by
obscuration, but not a correct interpretation in our case. Here, most
of the selected smaller and fainter galaxies are objects that are seen
through a thick extinction layer and hence are intrinsically large.
This can be visualized by comparing the top panels of
Fig.~\ref{allhist} with the bottom panels. The latter show the same
histograms but with the parameters corrected for extinction according
to Cameron's law (1990). The mean apparent magnitude of the total
sample is brighter by about 2 magnitudes (from $<B_{\rm J}> = 16\fm2$ to
$<B_{\rm J}^{\rm o}> = 14\fm3$) and the diameters larger by a factor of nearly 2
(from $<D> = 55\arcsec$ to $<D^{\rm o}> = 103\arcsec$), hence a sample of
the intrinsically brightest and largest spiral galaxies in the ZOA. As
the \HI\ emission is not hindered by dust, detections of these objects
should be equally likely.

\begin{figure}[t]
\hfil \epsfig{file=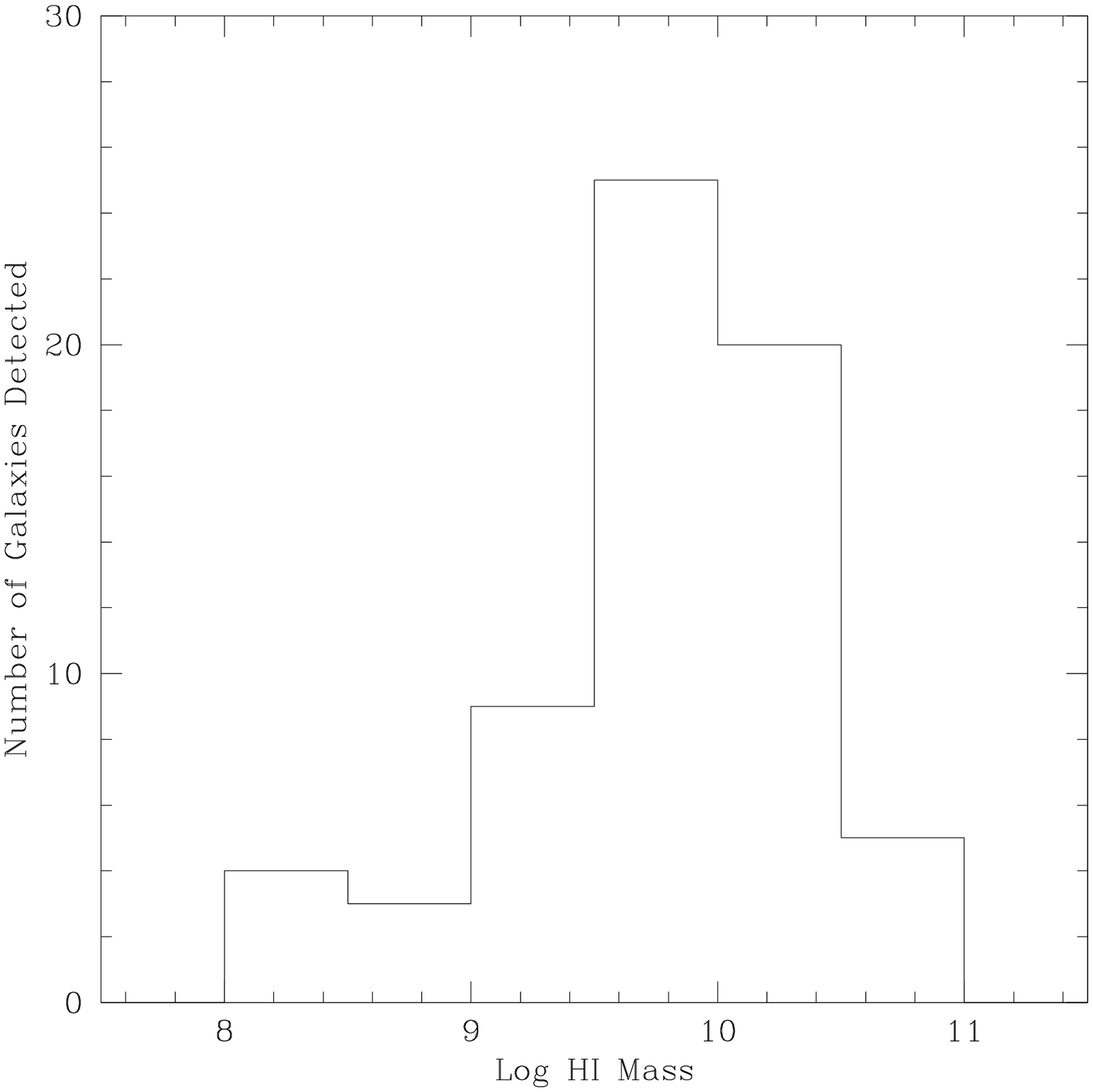,width=7.95cm} \hfil
\caption{Distribution of the \HI\ masses of the 66 detected objects.}
\label{masshist}
\end{figure}

It might thus seem a bit surprising that a fair fraction ($\sim 40\%$)
of the brighter ($B_{\rm J}^{\rm o} \la 14\fm0$) and extended ($D^{\rm o} \ga
90\arcsec$, $\log D^{\rm o} \ga 1.9$) observed spiral galaxy sample was not
detected in \HI\ (see bottom panels). A closer inspection of the
brightest and largest of these non-detections reveals that they stem
from pointings to extremely obscured objects at high Galactic
extinctions ($A_{\rm B} \ge 2\fm5$). The Cameron corrections to the true
apparent sizes become very large (e.g. a factor of $f \ge 4.5$ for the
diameters for $A_{\rm B} \ge 3\fm0$) and increasingly uncertain.  Moreover,
with such heavy absorption, the morphological classifications of these
objects remains guesswork. Some of these objects might not be spiral
galaxies -- or galaxies at all. Taking this into account,
i.e. ignoring the few heavily obscured, possibly misclassified objects
at extreme extinctions, our interpretation of Fig.~\ref{allhist} is
that the \HI\ observations yield a high detection rate and are very
powerful in tracing the nearby bright spiral galaxy population across
the Milky Way.  Indeed, the average \HI\ mass of the detected sample,
$9 \cdot 10^9\msun$, and the distribution of \HI\ masses of the
detected objects, shown in Fig.~\ref{masshist}, indicate the detected
galaxies are mostly normal spirals, with a smaller number of lower
\HI\ mass objects.

\section{Large-scale structure in the Hydra/Antlia-region} \label{lss}

Figure~\ref{veldist} shows a histogram of the radial velocities
(heliocentric) of the 66 \HI\ detections. The distribution indicates
that our sensitivity of typically rms $2 - 4$\,m\,Jy results in a fair
coverage of the galaxy distribution out to redshifts of about $V \sim
6000$\kms\ but does not probe the Universe well beyond that
distance. The distribution does not match the expectation for a
homogenously filled volume. Due to the different structures present in
this volume (see discussion below) it is very flat in the velocity
range $2000 \la V \la 6000$\kms\ except for a distinct peak at $\sim
3000$\kms. The latter is the signature from the 
extension of the Hydra/Antlia filament on the opposite side of the
Galactic Plane, which we confirm with this work.

\begin{figure}[t]
\epsfig{file=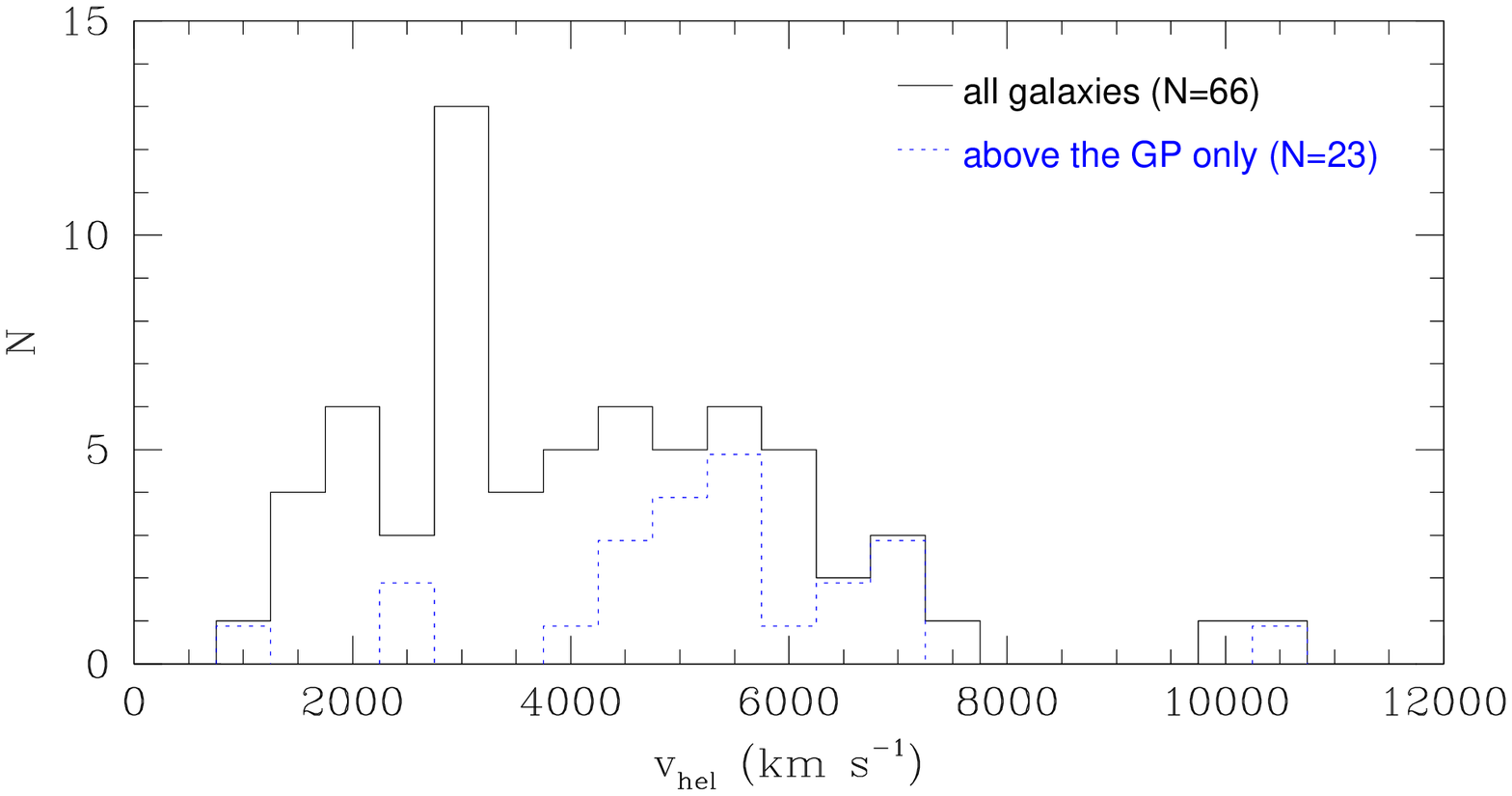,width=8.8cm} 
\caption{Velocity distribution of the 66 \HI\ detections. Good
sensitivity is achieved out to 6000\kms. Note the distinct peak
centered at 3000\kms\ which is due to galaxies below the GP only,
as can be verified when comparing all detections (solid line) with
detections below the GP (dotted line) which mainly 
finds galaxies in the range $4000 - 7000$\kms.}
\label{veldist}
\end{figure}

\begin{figure*}[t]
\epsfig{file=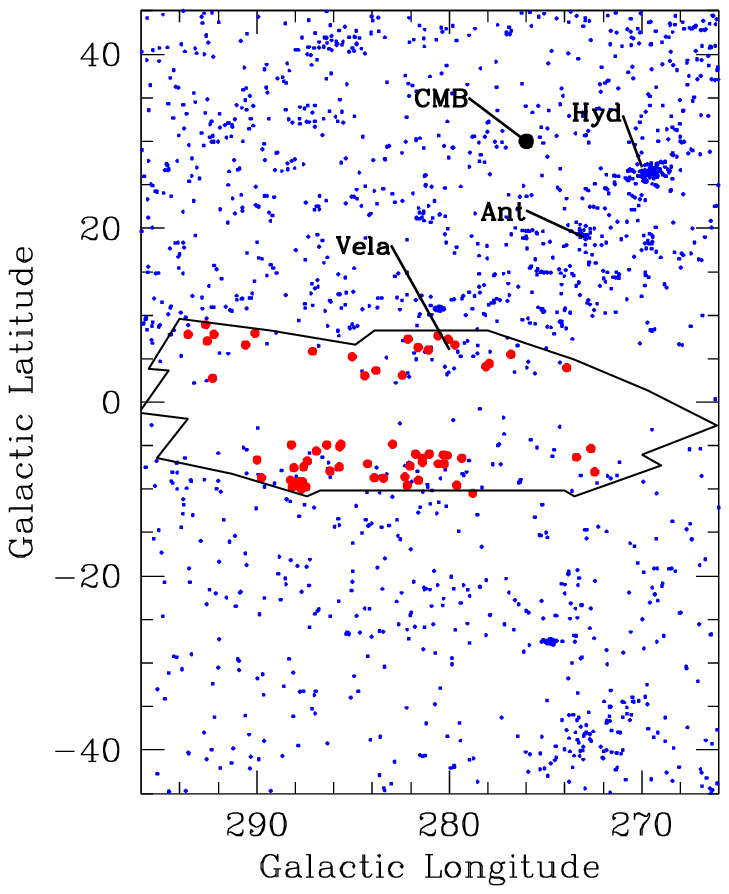,height=10.7cm,width=8cm,bbllx=39pt,bblly=330pt,bburx=249pt,bbury=588pt,clip=,angle=0} \hspace{0.5cm}
\epsfig{file=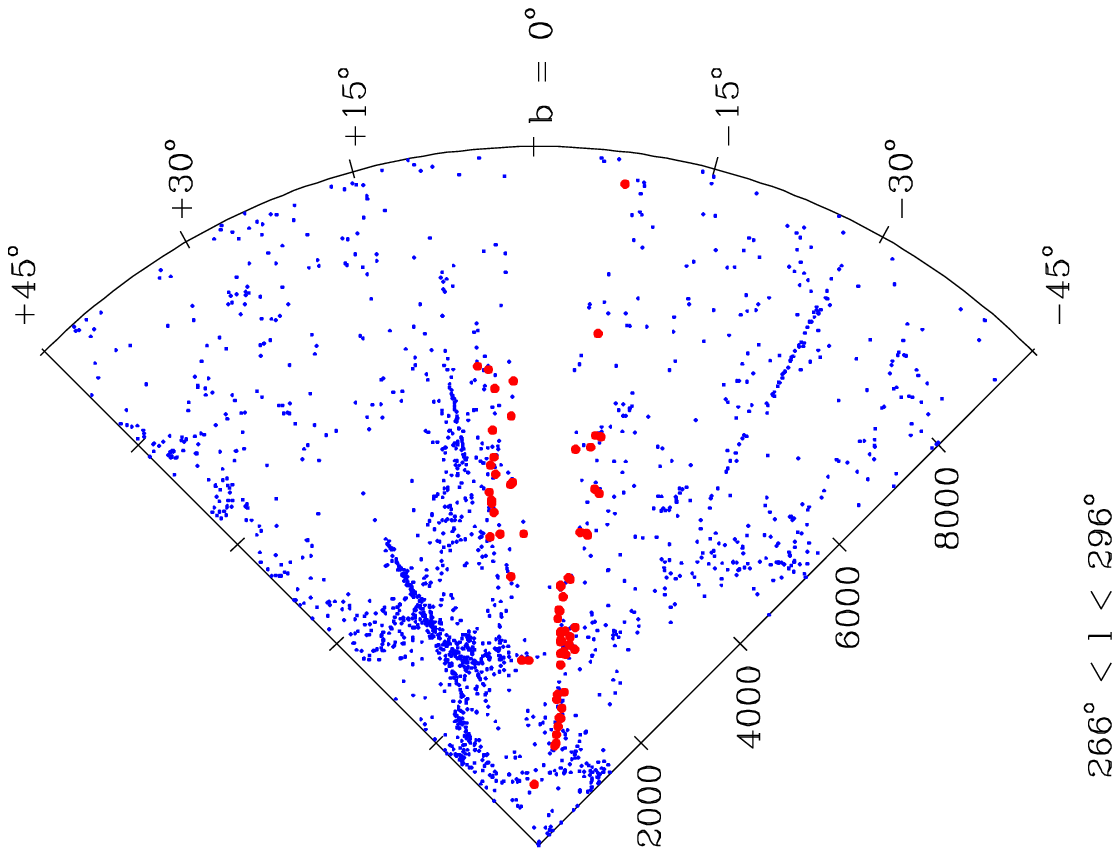,width=11.2cm,bbllx=585pt,bblly=365pt,bburx=257pt,bbury=614pt,clip=,angle=90}
    \caption{Distribution of all galaxies with $V \le 10\,000$\kms\
within $\pm 45\deg$ of the GP centered on the Hydra/Antlia search
area.  Previously measured galaxies are marked with small dots, the newly
obtained \HI\ galaxies with large filled circles. The left panel shows a sky
projection in Galactic coordinates in which the Hydra and Antlia
clusters, the Vela supercluster and the CMB dipole are also marked.  The right
panel shows a redshift cone out to $V \le 10\,000$\kms\ of the same
galaxies and with the same latitude orientation.}
\label{slice}
\end{figure*}

To visualize how the new detections add to our comprehension of LSS in
this region of the sky, we have combined the new data with other data
in and surrounding this area. Fig.~\ref{slice} displays galaxies
extracted from the LEDA database with $V \le 10\,000$\kms\ and $|b|
\le 45\deg$ (small dots) centered on the galaxy search area including
the new \HI\ detections as well (large filled circles). The left panel
shows the resulting sky projection in Galactic coordinates. The main
features in this region and volume, the Hydra and Antlia clusters, the
Vela supercluster and the dipole direction of the Cosmic Microwave
Background (CMB) radiation (Kogut \etal 1993), are labelled for
orientation. It is obvious that the new data add considerably to the
velocity data in the ZOA even though the LEDA database already
includes the optical spectroscopy obtained by us in the search area.

Studying the new data and its distribution in space in more detail, we
find that the clumpiness of the \HI\ detections seen in the sky
projection is due to four main distinct LSS features, whereas the 
mainly empty regions are due to galaxies which belong to structures 
that cross the GP at higher redshifts, -- evident as the high rate
of non-detections in those regions.

Astonishingly, only 2 of the detections above the GP around
$\ell \sim 280\deg$ form part of the suspected filament from the Hydra
and Antlia clusters across the ZOA. The other detections all lie
within 5000\kms\ to 7000\kms\ and thus are members of the Vela
Supercluster.

The galaxies in the upper left-hand corner of the search area lie
within a very narrow velocity range ($4500 - 5500$\kms). They 
belong to the Norma Supercluster (or Great Attractor overdensity)
and lie on the edge of this very broad extended wall-like structure 
which has the Norma cluster (ACO 3627) at its center, and crosses the
GP at fairly flat angle towards the Vela Supercluster (Fairall \etal  
1998).

The galaxies below the GP with velocities $1400 - 1900$\kms\ are
located in the galaxy concentration at $\ell \sim 287\fdg5, b \sim
-9\fdg5$ (see left panel). The redshift cone indicates that they form
part of the surface which outlines a void centered at $\sim 2000$\kms\
and $+10\deg$ latitude.

Although only 2 galaxies were detected above the GP at $\ell \sim
280\deg$ that form part of the Hydra/Antlia extension, practically all
the galaxies below the plane at these longitudes are found within the
narrow velocity range of $2100 - 3100$\kms. These data therefore seem
to support the existence of a filamentary structure that stretches
from the Hydra cluster at ($\ell=270\deg, b=28\deg, V\sim3300$\kms) to
the Antlia cluster ($273\deg, 19\deg, \sim2800$\kms), continues across
the GP (this bridge is confirmed in the preliminary analysis of the
Parkes deep ZOA \HI\ survey, Staveley-Smith, priv. comm.) and emerges on
the opposite side as an agglomeration of galaxies at about
($280\deg, -7\deg, \sim2500$\kms). Whether it stops there -- marking yet
another boundary of the previously mentioned void -- or continues even
further along the surface on another bubble (centered at 3500\kms\ and
$b \sim -45\deg$) is uncertain.
   
The newly revealed and substantiated structures seen in
Fig.~\ref{slice} confirm that \HI\ observations of highly obscured
spiral galaxies are an important tool in mapping large-scale
structures hidden by the Milky Way since the \HI\ observations probe
deeper into the ZOA than do optical spectroscopic observations.

\section{Summary}
The pointed 21-cm line observations of partially- to heavily-obscured
spiral galaxies uncovered in the deep optical search for galaxies in
the ZOA has proven this technique to be very powerful for
mapping a population of galaxies -- which on average actually are
intrinsically large and bright -- whose redshifts would be 
difficult or impossible
to obtain otherwise.
 
Besides the chance detection of a few galaxies without optical
counterparts (see Sect.~\ref{weirdt}), the 66 detections -- and the
distribution of the non-detections -- have helped to delimit a number
of suspected or previously unknown large-scale structures, in
particular the continuation of the Hydra/Antlia bridge across the GP
and the void with a radius of about 1000\kms\ centered in the
ZOA at about 2000\kms.

These HI-observations helped reduce the width of the ZOA in
redshift space. They will also be invaluable in mapping the peculiar
velocity field in the ZOA. The missing gap in LSS ($|b| \la 5\deg$)
will be filled with the detections from the deep ZOA survey
that is being carried out, at similar sensitivities, 
with the multibeam receiver on the Parkes radiotelescope.

\begin{acknowledgements}
It is with pleasure that we thank the ATNF staff at the Parkes
telescope for their high quality and cheerful support during our
observing runs.  Part of this survey was performed at the Kapteyn
Astronomical Institute of the University of Groningen and at the
Observatoire de Paris-Meudon. Their support is greatfully
acknowledged.  RCKK thanks CONACyT for their support (research grant
27602E). During much of this work, the research of PH was supported
by NSF Faculty Early Career Development (CAREER) Program award AST
95-02268. We have also made use of the Lyon-Meudon Extragalactic
Database (LEDA), supplied by the LEDA team at the Centre de Recherche
Astronomique de Lyon, Observatoire de Lyon, and of the NASA/IPAC
Extragalactic Database (NED), which is operated by the Jet Propulsion
Laboratory, Caltech, under contract with the National Aeronautics and
Space Administration.

\end{acknowledgements}

\end{document}